\documentclass[aip,preprint,pof]{revtex4-1}
\usepackage{CJK}
\usepackage[francais,english]{babel}
\usepackage[utf8]{inputenc}
\usepackage[T1]{fontenc}
\usepackage[pdftex]{graphicx}
\usepackage{setspace}
\usepackage{hyperref}
\usepackage[french]{varioref}
\usepackage{caption}
\usepackage{amsmath,amsfonts,amssymb}

\begin{document}

\title{On the front shape of an inertial granular flow down a rough incline}
\date{\today}

\author{G. Saingier}
\affiliation{Sorbonne Universit\'es, UPMC Univ Paris 06, CNRS UMR 7190, Institut Jean le Rond d'Alembert, F-75005 Paris, France}
\affiliation{Surface du Verre et Interfaces, UMR 125, CNRS/Saint-Gobain, 93303 Aubervilliers, France}
\author{S. Deboeuf \footnote{Author to whom correspondence should be addressed.  Electronic mail:  sdeboeuf@dalembert.upmc.fr } } 
\affiliation{Sorbonne Universit\'es, UPMC Univ Paris 06, CNRS UMR 7190, Institut Jean le Rond d'Alembert, F-75005 Paris, France}
\author{P.-Y. Lagr\'ee}
\affiliation{Sorbonne Universit\'es, UPMC Univ Paris 06, CNRS UMR 7190, Institut Jean le Rond d'Alembert, F-75005 Paris, France}

\begin{abstract}

Granular material flowing on complex topographies are ubiquitous in industrial and geophysical situations. In this paper, we study the small-scale model of a granular layer flowing on a rough incline. The shape of a granular front is solved analytically by using a 1D Savage-Hutter's model based on depth-averaged mass and momentum equations with the fractional expression for the frictional rheology $\mu$($I$). 
Unlike 
previous studies where a "plug flow" is assumed, a free shape factor $\alpha$ describing the vertical velocity profile, is taken into account to determine the solution. Such a way, we put in evidence an effect of inertia through the Froude number $Fr$ and the shape factor $\alpha$ on the front profile. The analytical predictions are compared with experimental results published by [O. Pouliquen, Phys. Fluids \textbf{11}, 1956 (1999)] and with our new experimental data obtained at higher Froude numbers. A good agreement between theory and experiments is found when $\alpha = 5/4$ is used in our model, corresponding to a Bagnold-like velocity profile. However, open questions are raised about the vertical velocity profile in granular flows and about the expression of the rheological function $\mu$($I$) and  its calibration from experimental data. 

\end{abstract}

\pacs{?}

\maketitle


\section{Introduction}

The flow of granular material on inclined topographies is a fundamental situation encountered in many industrial applications (chemical engineering, civil engineering, food-processing industry) and geophysical situations (rock avalanches, pyroclastic flows). This situation has aroused  extensive experimental, numerical and theoretical works based on model systems \cite{GDR_MiDi_2004, Duran_2012, Andreotti_2013, Delannay_2015} for several decades. In spite of these numerous studies, no constitutive laws are currently able to predict and explain all the range of behaviours of a dry cohesionless granular material \cite{Forterre&Pouliquen_2008}.

The first  system of closed equations for a granular flow was proposed by Savage \& Hutter \cite{Savage_1989} in 1989 by depth-averaging the mass and momentum equations, using a constant Coulomb basal friction law. This theoretical model looks like the Saint-Venant shallow-water equations - commonly used for liquids - with an additional source term. This approach needs to do some hypothesis on the shape of the velocity profile in the depth by determining the value of the shape factor $\alpha$, defined later in formula (\ref{alpha}). Many authors choose to consider a "plug flow" profile in order to simplify the equations. The same problem appears for newtonian shallow water flows where the influence of the shape factor is often eluded. Nevertheless, Hogg \& Pritchard \cite{Hogg&Pritchard_2004} have put in evidence the importance of this shape factor to correctly describe the inertial flows of viscous laminar fluids.

In 1999, Pouliquen \cite{Pouliquen_1999b} used the Savage \& Hutter's model to explain his experimental results of granular front profiles of a steady uniform flow on an inclined plane. He uses an empirical basal friction \cite{Pouliquen_1999a} instead of a constant friction. Different expressions for  this friction law are  proposed in the literature\cite{Pouliquen_2002, DaCruz_2005, Jop_2006, Hatano_2007}. Following these works, the local $\mu$($I$)-rheology has recently emerged as an appropriate framework to describe experimental observations, discrete numerical simulations and to compute continuous numerical simulations \cite{GDR_MiDi_2004, DaCruz_2005, Jop_2005, Lagree_2011, Staron_2014}. 
In a simple shear flow of grains of diameter $d$ and density $\rho$, this formalism describes the friction coefficient $\mu$, corresponding to the ratio of the shear stress $\tau$ and the normal stress $P$, as a function of the  inertia number $I$, depending on the pressure $P$ and the shear rate $\dot{\gamma}$, defined as:
\begin{equation}
I = \frac{\dot{\gamma} d}{\sqrt{P/\rho}}.
\end{equation}

In this paper, we propose a new analytical solution for the granular front of a steady uniform flow on an inclined plane by using the Savage \& Hutter's model with the fractional $\mu$($I$)-rheology as defined in Jop \textit{et al.} \cite{Jop_2006}. By taking into account the shape of velocity profile, we will show that the front profile depends on the velocity profile and  the Froude number. This prediction is confirmed by a comparison with  new experimental results of granular flows on a rough inclined plane at high Froude numbers. 

This paper begins in Sec. II by the introduction of the theoretical model and the resolution of the analytical front profile. In Sec. III, the experimental set-up, the measurement methods and the first experimental observations are presented. The comparison between experimental data and theoretical predictions is done in Sec. IV by using results from Pouliquen \cite{Pouliquen_1999b} and our new experimental results at higher Froude numbers. Our  results are discussed in Sec. V. 

\section{Analytical solution for the front profile}

We consider a thin layer, transversally uniform, of a granular material of solid fraction $\phi$
composed of grains of diameter $d$ and density $\rho$. We assume that the
granular flow is incompressible and we will take the solid fraction equal to $\phi = 0.6$. The granular material flows over a rough inclined surface, that is assumed to impose a no-slip condition at the bottom. The streamwise and vertical coordinates are denoted by $x$ and $z$, and $h(x,t)$ denotes the depth of the layer. The slenderness of the granular layer allows us to use the shallow-water Saint-Venant equations in 1D written by Savage \& Hutter \cite{Savage_1989}:  
\begin{equation}
\frac{\partial h}{\partial t} + \frac{\partial }{\partial x} (h\overline{u}) = 0,
\label{St_Venant_masse}
\end{equation}
\begin{equation}
\frac{\partial }{\partial t} (h\overline{u}) + \alpha \frac{\partial }{\partial x} (h\overline{u}^{2}) = hg\cos\theta (\tan\theta - \mu(\overline{I}) - \frac{\partial h}{\partial x}),
\label{St_Venant_qdm}
\end{equation}
where $\overline{u}$ denotes the depth-averaged velocity. The first term of the right hand side is the gravity along the slope, the second is the basal friction and the third is the pressure gradient. Note that the earth pressure coefficient $k$ is taken equal to $1$ -- this describes the redistribution of normal stresses\cite{Wieghardt_1975, Pouliquen_1999b} --. 
Recently  these equations have been  revisited by Gray \& Edwards \cite{Gray_2014} by integrating Navier-Stokes equations with a local $\mu$($I$)-rheology. 
We introduce the shape factor $\alpha$ usually defined as
\begin{equation}
 \alpha = \frac{\frac{1}{h}\int_0^h u^2(z)dz}{\left( \frac{1}{h} \int_0^h u(z)dz \right)^2}.
\label{alpha}
\end{equation}
In many papers, the simplification $\alpha = 1$ is carried out by the authors (Savage \& Hutter \cite{Savage_1989}, Iverson \textit{et al.} \cite{Iverson_1997}, Pouliquen \cite{Pouliquen_1999b}, Mangeney \textit{et al.} \cite{Mangeney_2003}, Gray \& Edwards \cite{Gray_2014}). This simplification implies that the material presents a uniform velocity profile in the vertical direction. The material flows like a solid without shear ("plug flow"). This representation may be really inappropriate to describe the flow of a granular thin layer regarding the Bagnold-like profile for the velocity (see GDR MiDi \cite{GDR_MiDi_2004}) defined by 
\begin{equation}
\frac{u(z)}{\sqrt{g d}} = \frac{2}{3} \overline{I} \sqrt{\cos\theta} \frac{(h^{3/2} - (h-z)^{3/2})}{{d}^{3/2}}.
\label{champ_vitesse}
\end{equation}
For this velocity profile, the mean velocity $\overline{u}$ and the mean inertial number $\overline{I}$ are defined respectively by
\begin{equation}
\overline{u} = \frac{3}{5} u(h) \ \ \ \text{and} \ \ \ \overline{I} = \frac{5}{2} 
\frac{\overline{u} d}{h \sqrt{\phi g h \cos\theta}},
\label{defI}
\end{equation}
where $u(h)$ is the free surface velocity. With this expression for the velocity profile, the calculation of the shape factor leads to $\alpha = 5/4$. In this paper, we do not consider the usual simplification $\alpha = 1$ and we will discuss the effect of the $\alpha$ value.
The friction $\mu$($I$) is expressed here with the fractional friction law proposed by Jop \textit{et al.} \cite{Jop_2006}:
\begin{equation}
\mu(I) = \mu_{0} + \frac{\Delta\mu}{I_{0}/I + 1},
\label{rheol_frac}
\end{equation}
where $\mu_0$, $\Delta\mu$ and $I_0$ are empirical parameters characterizing the granular set-up.

With (\ref{St_Venant_masse}), (\ref{St_Venant_qdm}), (\ref{rheol_frac}) and appropriate boundary conditions, we can solve the problem for any shallow granular flow. In order to derive the analytical front profile of an uniform flow, we have to solve this system of equations in the case of the front propagation, with the boundary condition $h =h_\infty$ = cst far upstream to the front. As observed experimentally by Pouliquen \cite{Pouliquen_1999b} (and as we will show in the next part, see Fig. \ref{front_propagation}), the front moves at a constant velocity $u_0$ without deformation, leading to a travelling wave for the front propagation:
\begin{equation}
h(x,t) = h(\xi) \ \ \text{with} \ \ \xi = x - u_{0}t.
\end{equation}
The mass balance equation (\ref{St_Venant_masse}) becomes $d(h(\overline{u} - u_0))/d \xi = 0$,  implying that $\overline{u} = u_0$. 
Far upstream to the front, the flow tends toward a steady uniform flow characterized by the thickness $h_\infty$ and the velocity $u_0$. By using this change of variables ($\xi = x - u_{0}t$) and by introducing the Froude number $Fr$, associated to the steady uniform flow: 
\begin{equation}
Fr = \frac{u_{0}}{\sqrt{g h_\infty \cos\theta}},
\label{Froude}
\end{equation}
the depth-averaged momentum balance equation (\ref{St_Venant_qdm}) can be rewritten as in the moving frame:
\begin{equation}
\left[ (\alpha - 1)Fr^{2}\frac{h_\infty}{h} + 1 \right] \frac{d h}{d \xi} = \tan\theta -\mu(\overline{I}).
\label{eq_gene_Fr}
\end{equation}
Since the depth-averaged velocity $\overline{u}$ is the same everywhere, equal to $u_0$, it is possible to determine it by using the  Bagnold-like velocity profile defined previously. In each point of the front, the velocity $u_0$ is
\begin{equation}
u_0 = \frac{2 \overline{I}_{\theta}}{5} \sqrt{\phi g h_\infty \cos\theta} \frac{h_\infty}{d} = \frac{2 \overline{I}}{5} \sqrt{\phi g h \cos\theta} \frac{h}{d},
\label{u_0}
\end{equation}
where $\overline{I}$ and $\overline{I}_\theta$ are the inertial numbers associated to the flow of thickness $h$ at the position $\xi$ and to the steady-uniform flow $h_\infty$ far upstream respectively.  
The equation (\ref{u_0}) leads to the relationship between $\overline{I}$ and $\overline{I}_\theta$:
\begin{equation}
\frac{\overline{I}_\theta}{\overline{I}} = (\frac{h}{h_\infty})^{3/2}.
\label{I_h}
\end{equation}
In the steady-uniform flow, the equation (\ref{eq_gene_Fr}) simplifies and $\theta$ can be expressed as a function of $\overline{I}_\theta$ by using the friction law: 
\begin{equation}
\tan\theta = \mu(\overline{I}_{\theta}) = \mu_{0} + \frac{\Delta\mu}{I_{0}/\overline{I}_{\theta} + 1}.
\label{tan_theta}
\end{equation}
The equations (\ref{I_h}) and (\ref{tan_theta}) allow us to replace $\bar{I}/I_0$ by a function of $h/h_\infty$:
\begin{equation}
\frac{\overline{I}}{I_0} = \left(\frac{h_\infty}{h}\right)^{3/2} \frac{\tan\theta - \mu_0}{\mu_0 + \Delta\mu -\tan\theta}.
\label{I_I0}
\end{equation}
By using the front equation (\ref{eq_gene_Fr}) with the frictional rheology (\ref{rheol_frac}), by introducing the relation (\ref{I_I0}) and by defining the non-dimensionalized variables: 
\begin{equation}
X = \frac{\xi(\tan\theta - \mu_0)}{h_\infty}, \ \ \ H = \frac{h}{h_\infty}, \ \ \  \delta = \frac{\tan\theta - \mu_0}{\Delta\mu},  
\label{adim}
\end{equation}
we obtain the non-dimensionalized equation for the front profile:
\begin{equation}
\frac{d X}{d H} = \frac{\delta + H^{3/2}(1-\delta)}{(\delta + H^{3/2}(1-\delta) -1 )(1 + \frac{(1-\alpha)Fr^2}{H})}.
\end{equation}
This equation has an implicit analytical solution $X(H)$ which can be expressed as:
\begin{multline}
X(H) = X_0 - \frac{1}{3(-1+\delta)}\times[ 3H(\delta-1) - 2\sqrt{3}\tan^{-1}(\frac{1+2\sqrt{H}}{\sqrt{3}}) - 2 \log(1-\sqrt{H}) \\
+ 3\delta(1-\alpha)Fr^2\log(H) + \log(1+\sqrt{H} + H) - 2(1-\alpha)Fr^2\log(1-H^{3/2})],
\label{sol_anal_impl}
\end{multline}
with $X_0$ an integration constant. So that:
\begin{equation}
\frac{h}{h_\infty} = X^{-1}\left[ \frac{x}{h_\infty} (\tan\theta - \mu_0) \right]. 
\end{equation}
Note that our analytical solution is different from the solution proposed by Pudasaini \cite{Pudasaini_2011} determined with the Bagnold's inertial stress \cite{Campbell_1990}. Our non-dimensionalized solution only depends on three parameters: $\delta$ accounting for the inclination and the rheology, $Fr$ for the inertia and $\alpha$ for the shape of the velocity profile. This solution presents an asymptotic exponential behaviour when $H$ tends to zero, for all $\alpha$ values excepted for $\alpha = 1$. Consequently, the analytical granular front for $\alpha \neq 1$ is preceded by a precursor film which may not be physical or not observed in experimental results. For comparisons with experimental data, the integration constant is imposed in order that the tangent to the inflection point crosses the origin point $(0,0)$.

\begin{figure}[h] 
\center
\includegraphics [width=80mm]{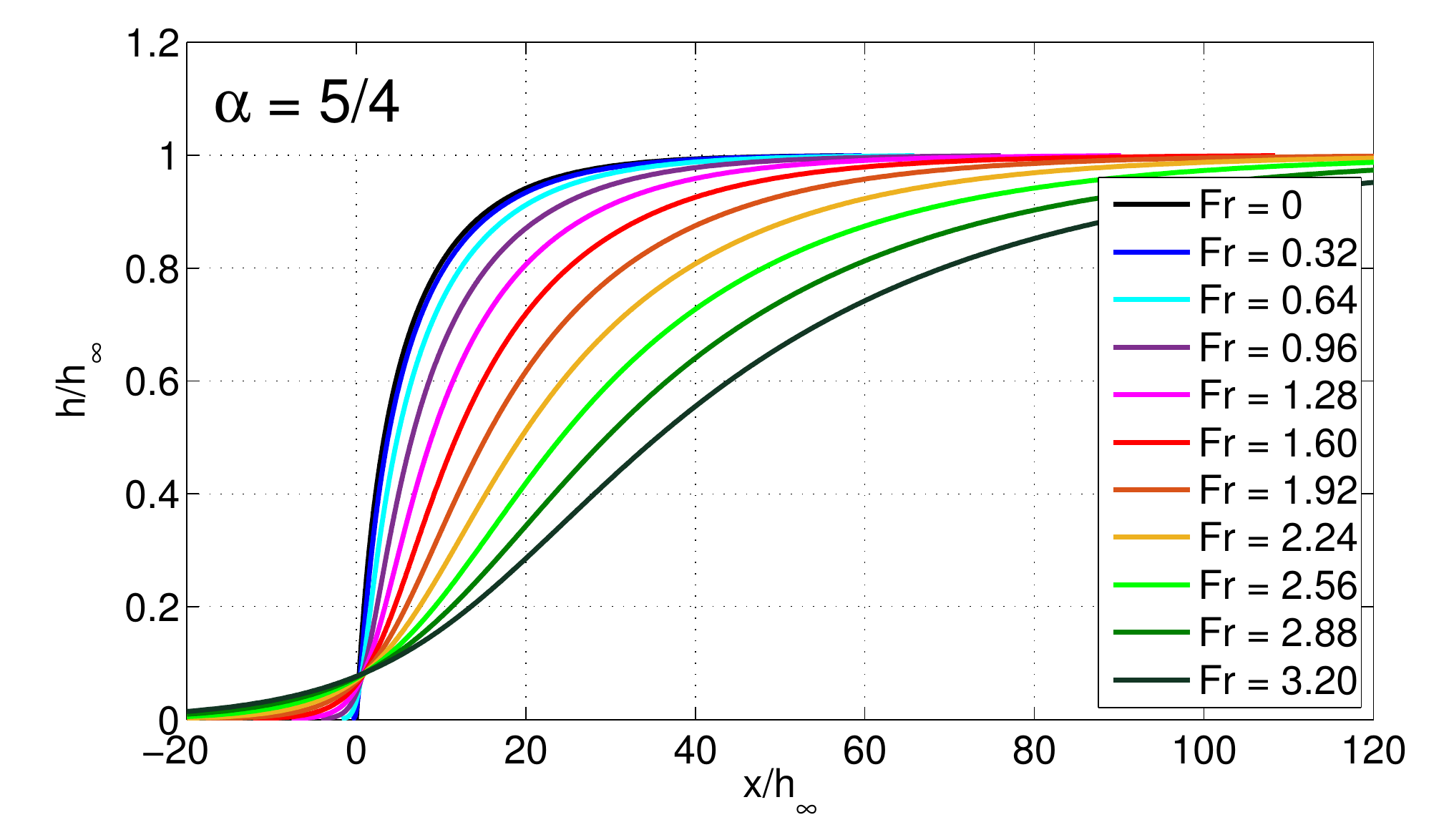} 
\includegraphics [width=80mm]{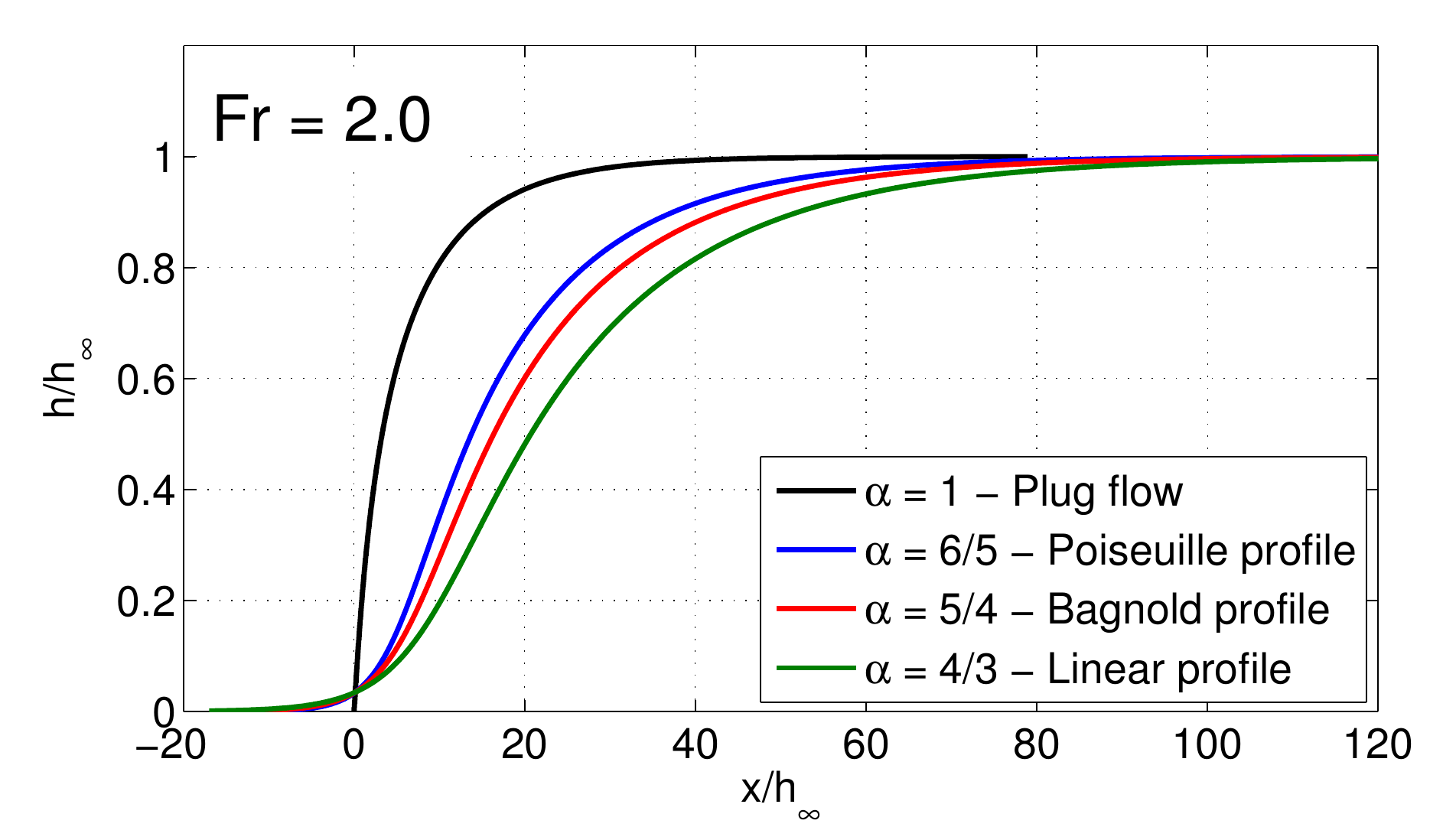} 
\captionof{figure}{Analytical solution for the front profile plotted for several sets of parameters at the inclination $\theta = 27^\circ$: (left) Effect of the Froude number $Fr$ for $\alpha = 5/4$; (right) Effect of the shape factor $\alpha$ for $Fr = 2.0$. }
    \label{sol_analytique}
\end{figure}

The effect of the different parameters on the front profile is discussed in Fig. \ref{sol_analytique}. Fig. \ref{sol_analytique} (a) shows several fronts for Froude numbers $Fr$ increasing from $0$ to $3.2$ for an inclination $\theta = 27^\circ$ and $\alpha = 5/4 = 1.25$ (Bagnold-like velocity profile). The non-dimensionalized front shape is flattened down by the inertial term. In Fig. \ref{sol_analytique} (b), front profiles are plotted at a constant $Fr=2.0$ for different values of $\alpha$ between $1$ and $4/3$. An increase of $\alpha$ also implies the flattening of the front. The "plug flow" profile corresponds to $\alpha = 1$. The case $\alpha = 5/4$ represents a Bagnold-like velocity profile whereas for $\alpha = 4/3$, the velocity profile is linear and for $\alpha = 6/5$ the profile corresponds to a Poiseuille profile.

Finally, considering the simplification $\alpha = 1$ implies to vanish all the terms which contain $Fr$. The analytical solution (\ref{sol_anal_impl}) can be reduced by choosing $\alpha = 1$ or $Fr = 0$, to the new solution:  
\begin{equation}
X(H) = \frac{(\delta-1) H-\frac{2}{3} \log \left(1-\sqrt{H}\right)+\frac{1}{3} \log \left(H+\sqrt{H}+1\right)-\frac{2 \tan ^{-1}\left(\frac{2
   \sqrt{H}+1}{\sqrt{3}}\right)}{\sqrt{3}}}{\delta-1}.
   \label{sol_anal_simpl}
\end{equation}
Note that in this case, the integration constant $X_0$ corresponds obviously to $H(X_0) = 0$. Consequently, the analytical solution for $\alpha = 1$ only depends on the inclination and the choice of rheology parameters. The precursor layer disappears and it is possible to measure a contact angle $\theta_c$ of the nondimensionalized profile between the granular fluid and the plane, which depends on the inclination and the rheology parameters:
\begin{equation}
\theta_c = \arctan(  (\mu_0 + \Delta\mu) - \tan\theta ). 
\end{equation}

\section{Experimental set-up}

\begin{figure}[t] 
\center
\includegraphics [width=100mm]{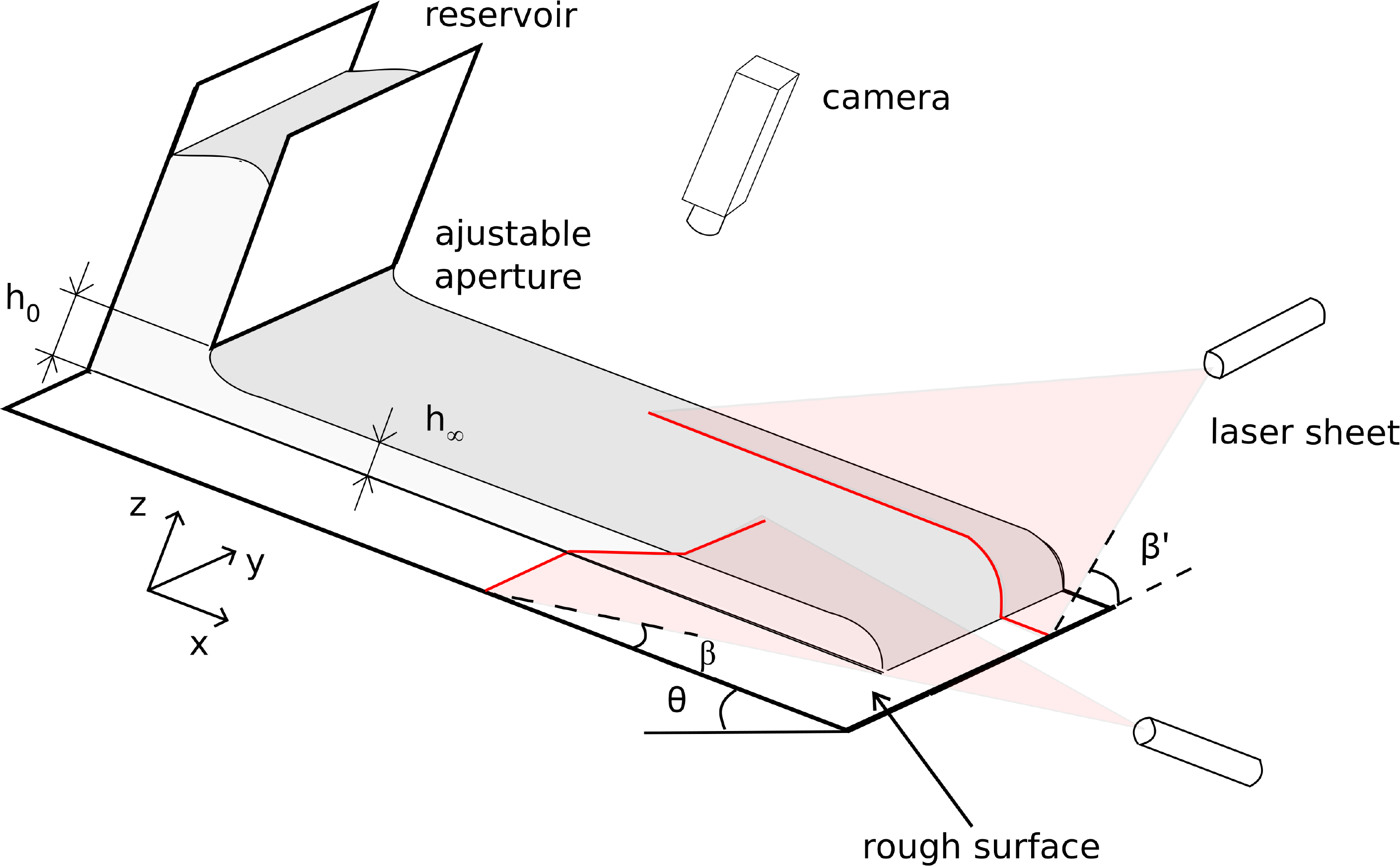} 
\includegraphics [width=60mm]{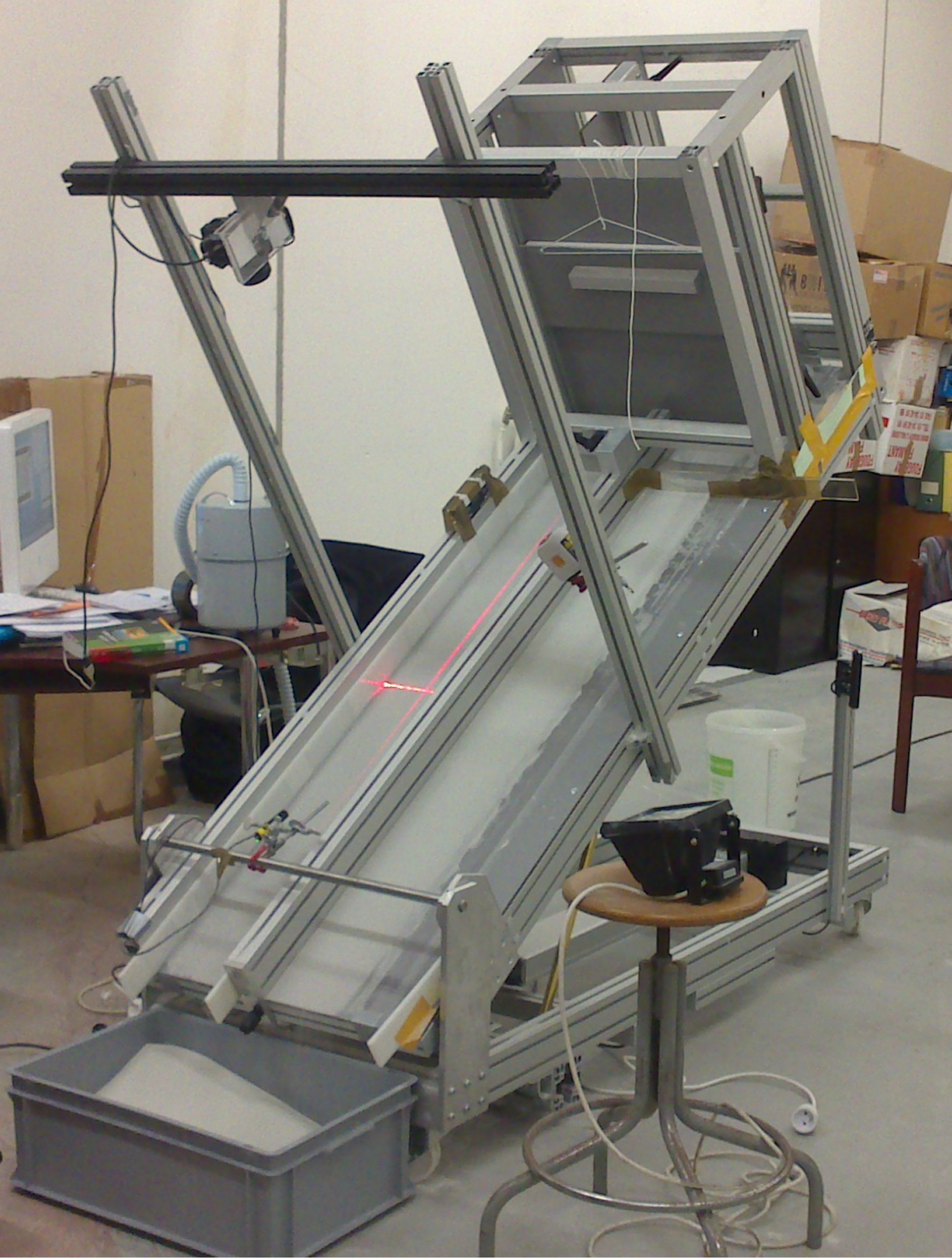} 
\captionof{figure}{Experimental set-up: (left) Schematic representation of the experimental set-up;  (right) Photograph of the set-up. }
    \label{schema}
\end{figure}

In order to check the theoretical predictions, we have revisited the experiments proposed by Pouliquen \cite{Pouliquen_1999b}. The propagation of a front of a dry granular material has been investigated experimentally thanks to classical experiments of inclined planes (see Pouliquen \cite{Pouliquen_1999a, Pouliquen_1999b}). The set-up, shown on Fig. \ref{schema}, is a 2-m-long and 40-cm-wide rough plane which can be inclined from $0^\circ$ to $32^\circ$. The granular material is stored in a reservoir at the top of the plane and is released through a gate which can be opened quickly and precisely. A second gate allows to adjust the aperture thickness in order to control the mass flow rate. The rough surface is obtained by gluing the same particles on the plane.

The granular material and the glued layer are composed of quasi monodispersed spherical glass beads of diameters $d = 200 \pm 50$ $\mu$m and the solid fraction is taken equal to $\phi=0.60$. The size of particles is  small enough in comparison with the size of the granular layer to justify the hydrodynamical continuous model used previously \cite{GDR_MiDi_2004}. Side walls are polyethylene plates to guarantee that the lateral conditions are smooth. In our experiments, only the centerline of the granular flow is studied to be assimilated to a 2D flow.

\begin{figure}[h] 
\center
\includegraphics [width=120mm]{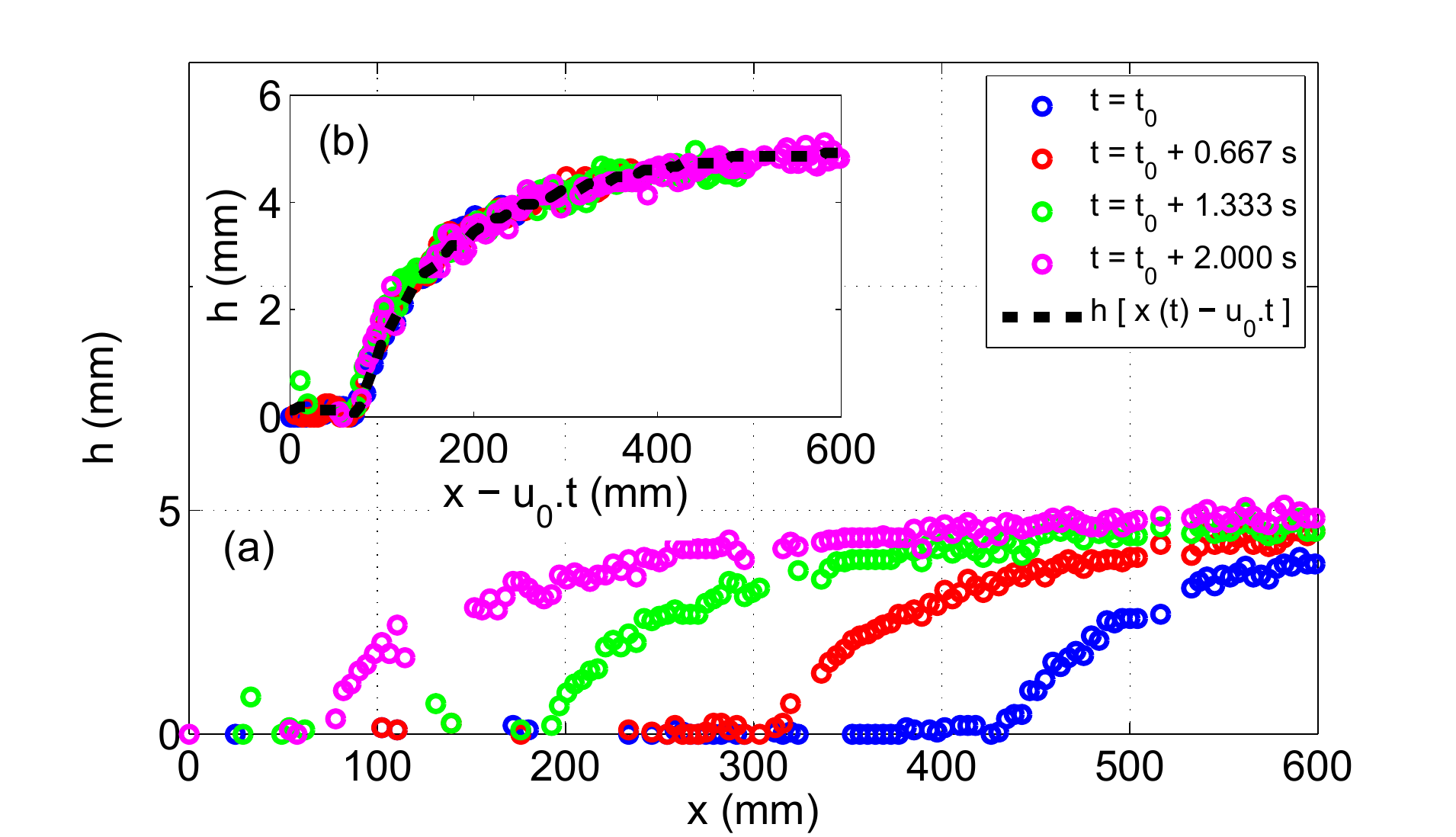} 
\captionof{figure}{(a) Granular front propagating at several times. (b) Superposition of the front profiles at different times by the change of variable $\xi = x - u_0 t$. The profiles obtained with spatial data are presented in filled circles, whereas the temporal front is shown with the dashed line after the variable change $t \rightarrow x=u_0t$. $\theta = 25.2^\circ$, $h_\infty = 4.9$ mm and $u_0 = 18$~cm/s. }
    \label{front_propagation}
\end{figure}

For the range of inclinations ($25^\circ$ to $30^\circ$) and aperture thicknesses ($5$ mm to $30$ mm) that we have studied, a granular front hurtles down the slope at constant velocity with a steady shape, as shown on Fig. \ref{front_propagation}.  The front velocity $u_0$ is measured by tracking the front propagating down the inclined plane with a home-made image processing algorithm. The thickness of the steady-uniform flow and the shape of the front are measured at a distance of $1$ m from the aperture in order to be unaffected by the transient region near the gate. 
The method of measurement consists to illuminate longitudinally the flow surface with a laser light sheet  at a low incident angle (see Pouliquen \cite{Pouliquen_1999a}). Where the granular flow crosses the projection of the laser sheet, it is shifted laterally from the initial position. The lateral shift is proportional to the thickness $h_t(x)$ and can be determined precisely after calibration. The spatial front profile at different times is represented in Fig. \ref{front_propagation} (a). It is straightforward that the front velocity is constant, when translating the front profiles by a constant velocity $u_0$ (see Fig. \ref{front_propagation} (b)). A second laser sheet illuminates the surface transversally with a smaller incident angle (see Deboeuf \textit{et al.} \cite{Deboeuf_2006}). 
Thus we obtain the transversal thickness at a position $x(t)$. By doing this measurement at several times, it is possible to determine a temporal evolution of the thickness $h_x(t)$. 
A comparison of both profiles ($h_t(x)$ and $h_x(t)$) is possible thanks to the change of variables $t \rightarrow x=u_0t$ or $x \rightarrow t=x/u_0$. 
As shown by the superposition of profiles on Fig. \ref{front_propagation} (b), this method of transversal profilometry leads to the same profile that the longitudinal profilometry, allowing for a higher resolution on a longer region of observation. It also proves that $\overline{u} = u_0$ in each position of the flow, showing that everything is actually constant in the moving frame. 

Let us now compare the experimental results with the theoretical ones by using previous experimental results\cite{Pouliquen_1999b} and new experimental data.

\section{Results}

\subsection{First case : Small Froude number}

First, we consider the case of slow granular flows ($Fr\simeq 0$). 
In this case, the inertial term can be neglected and the front equation (\ref{eq_gene_Fr}) simplifies to give the equation:

\begin{equation}
\frac{d h}{d \xi} = \tan\theta - \mu(\overline{I}).
\label{eq_simpl}
\end{equation}
The same equation is deduced if we consider $\alpha = 1$ as assumed in many papers \cite{Savage_1989, Pouliquen_1999b, Kerswell_2005, Gray_2014}, or more generally if $(\alpha-1)Fr^2 << 1$. Consequently, even if this simplification ($\alpha = 1$) is not physically justified  for granular Bagnold-like flows, it leads to a coherent equation for slow granular flows on inclines. 

Pouliquen \cite{Pouliquen_1999b} presented experimental results of granular material flowing on a rough plane. He observed a good collapse of experimental data of front profile $h(x)$ after rescaling by the steady-uniform thickness $h_\infty$. The equation (\ref{eq_simpl}) has been solved numerically by Pouliquen \cite{Pouliquen_1999b} with an exponential frictional rheology \cite{Pouliquen_1999a}: $\mu (I) = \mu_0 +\Delta\mu \exp(-I_0/I)$. More recently, Gray \& Edwards \cite{Gray_2014}
have proposed a numerical resolution with the fractional rheology (\ref{rheol_frac}). With the fractional rheology, the front profile is determined by the expression (\ref{sol_anal_simpl}) determined previously.

\begin{figure}[!b] 
\center
\includegraphics [width=80mm]{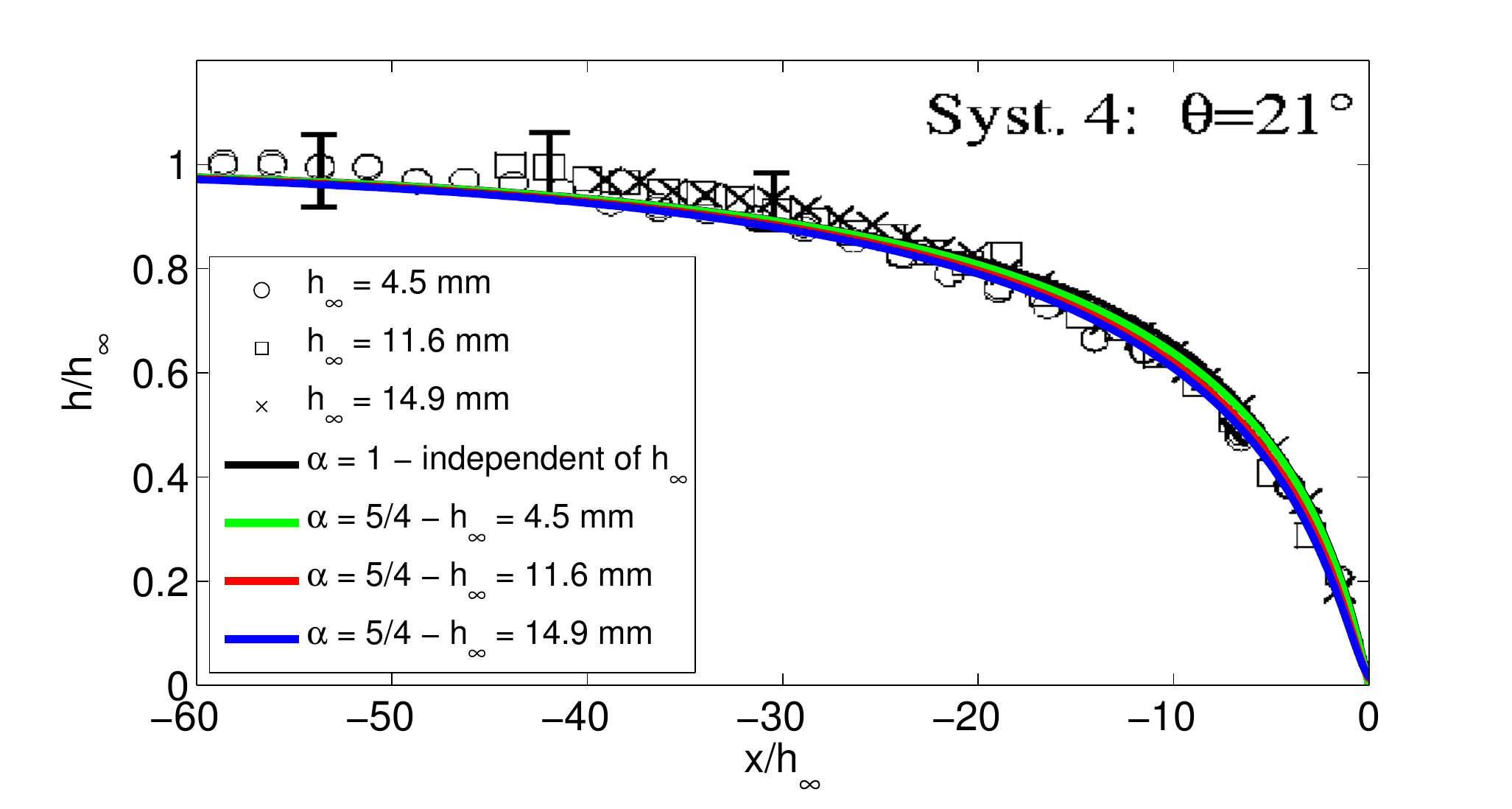} 
\includegraphics [width=80mm]{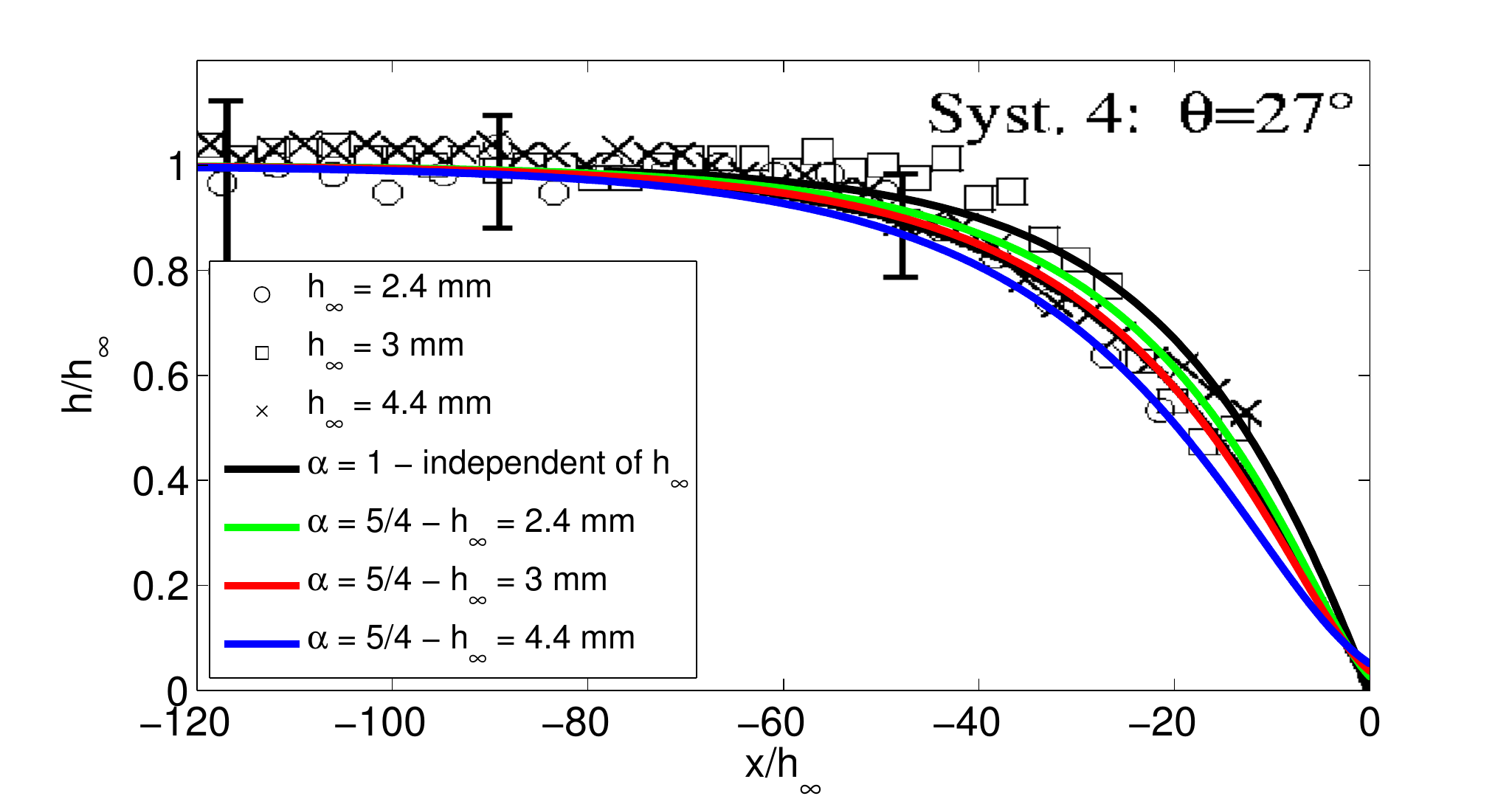} 
\captionof{figure}{Front profiles from Pouliquen \cite{Pouliquen_1999b}: comparison between experiments for different $h_\infty$ (symbols) and theory obtained with the fractional rheology (solid lines). Colored and dark lines correspond to the calculations with $\alpha = 5/4$ and $\alpha = 1$ respectively. The curves for $\alpha = 1$ with a fractional expression for the rheology are superimposed on the numerical fronts obtained by Pouliquen with an exponential expression. Rheology parameters are determined by fitting $h_{stop}$($\theta$) data from Pouliquen \cite{Pouliquen_1999a}: $\mu_0 = 0.35$ and $\Delta\mu = 0.21$.}
    \label{Pouliquen_syst4}
\end{figure}

In this analytical expression, the profile only depends on the parameter $\delta$ which depends on the inclination $\theta$ and the rheology parameters $\mu_0$ and $\Delta\mu$ (and not $I_0$). Consequently, for an imposed inclination $\theta$, the profiles are the same after dividing by $h_\infty$, as observed for the experimental data of Pouliquen\cite{Pouliquen_1999b}.

Fig. \ref{Pouliquen_syst4} is extracted from Pouliquen \cite{Pouliquen_1999b} and presents some experimental data obtained for one size of beads ($d=500\mu m$, system 4 \cite{Pouliquen_1999a}). The analytical solutions have been computed using a fit of $h_{stop}(\theta)$ data \cite{Pouliquen_1999a} associated to the fit of the flow rule for the characterization of the rheological parameters. The solution calculated for $\alpha = 1$ with the fractional form of the $\mu$($I$) rheology (Eq. \ref{sol_anal_simpl}) is superimposed on the numerical solution proposed by Pouliquen with the exponential form. Both of them well describe experimental data. Indeed the range of velocity of these granular flows (from 2 to 20 cm/s) corresponds to small Froude numbers since the typical thickness of the flow is $1$  cm (from $Fr = 0.1$ to $Fr = 1$). Consequently, the simplification leading to the equation (\ref{eq_simpl}) is relevant. By plotting the analytical solution calculated for $\alpha = 5/4$ (Eq. \ref{sol_anal_impl}) by using the velocity data \cite{Pouliquen_1999a}, we observe that the profiles are only slightly flattened but stay in the error bars of the data. 

\begin{figure}[b] 
\center
\includegraphics [width=80mm]{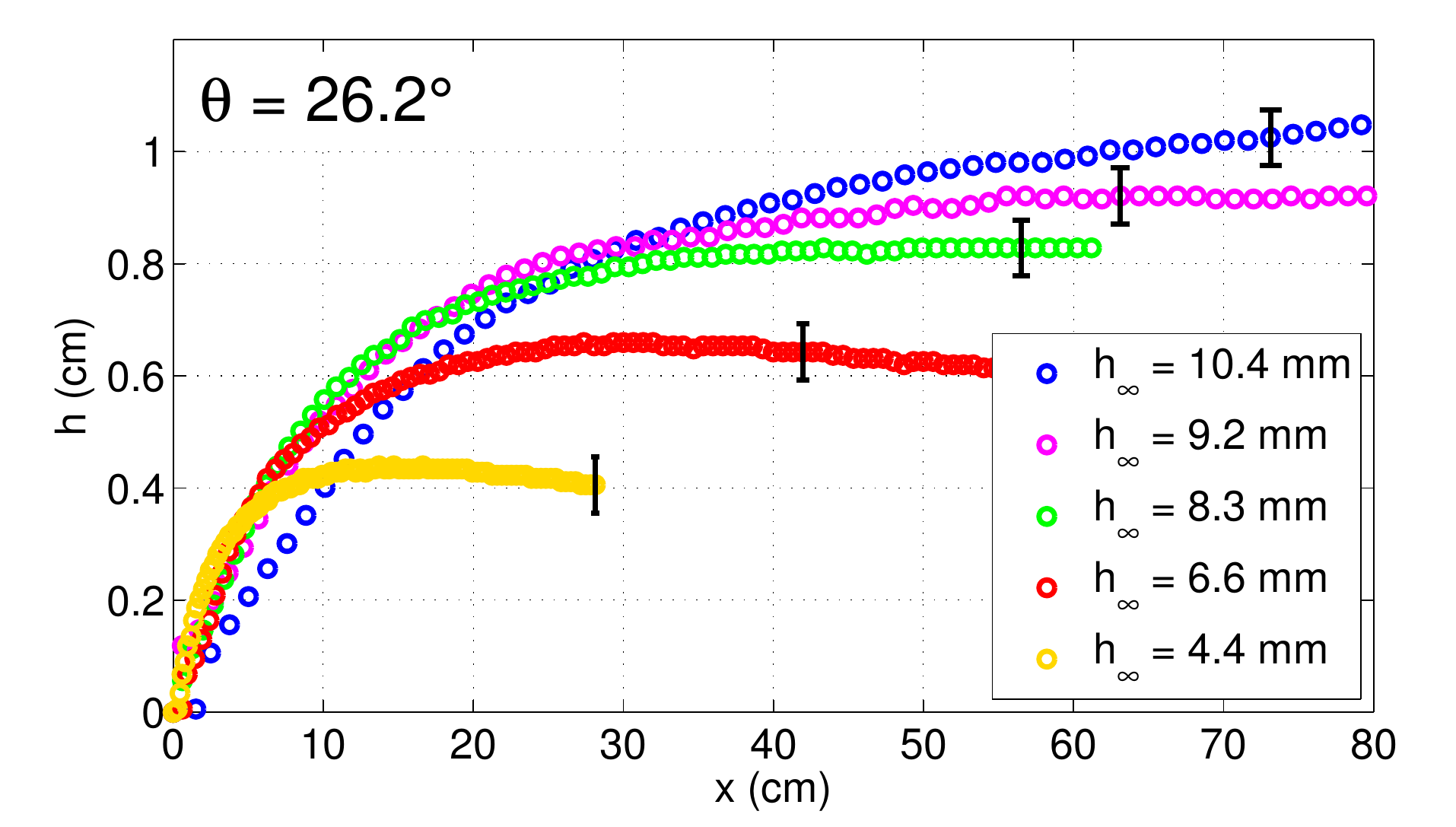} 
\includegraphics [width=80mm]{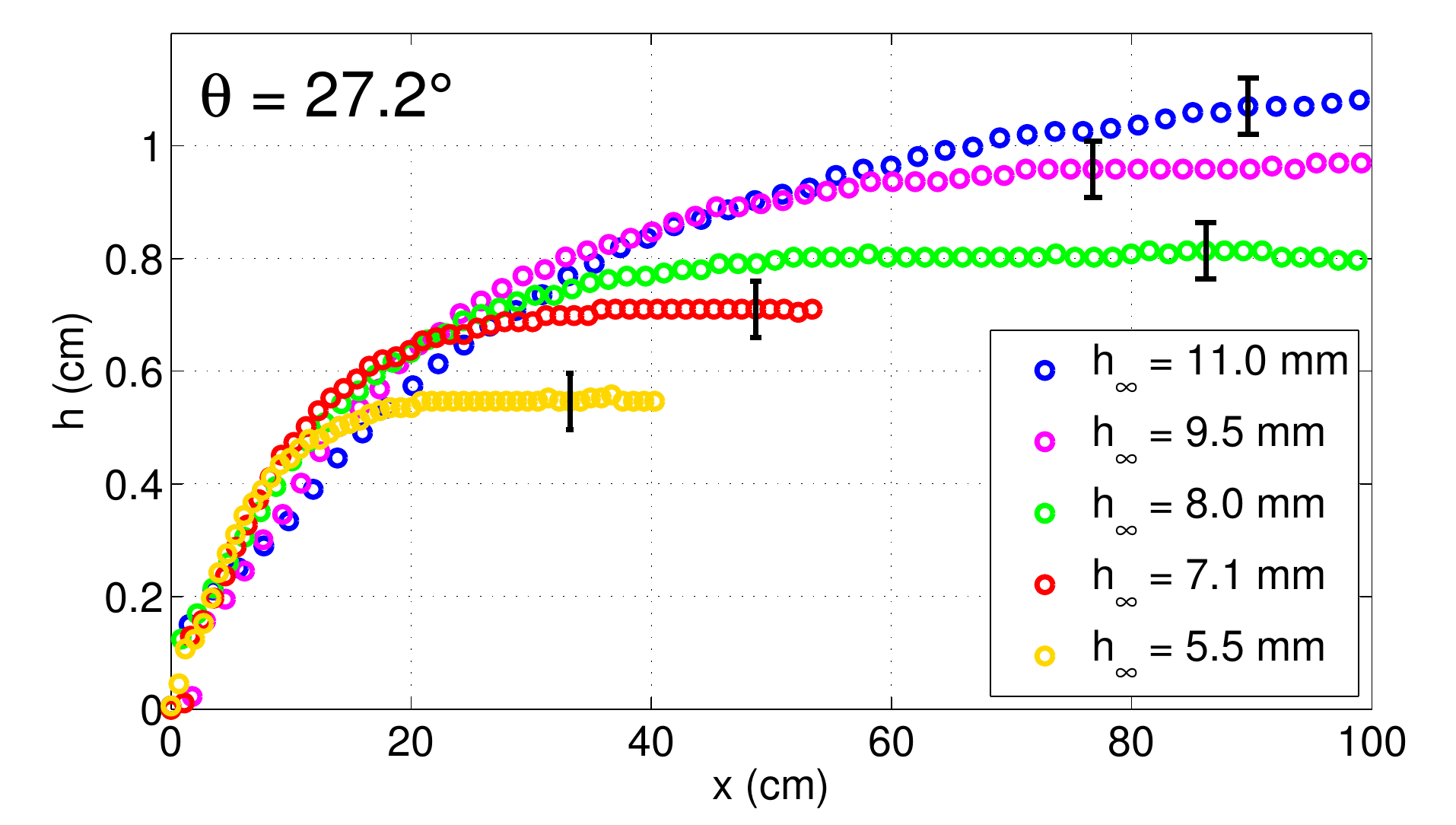} 
\includegraphics [width=80mm]{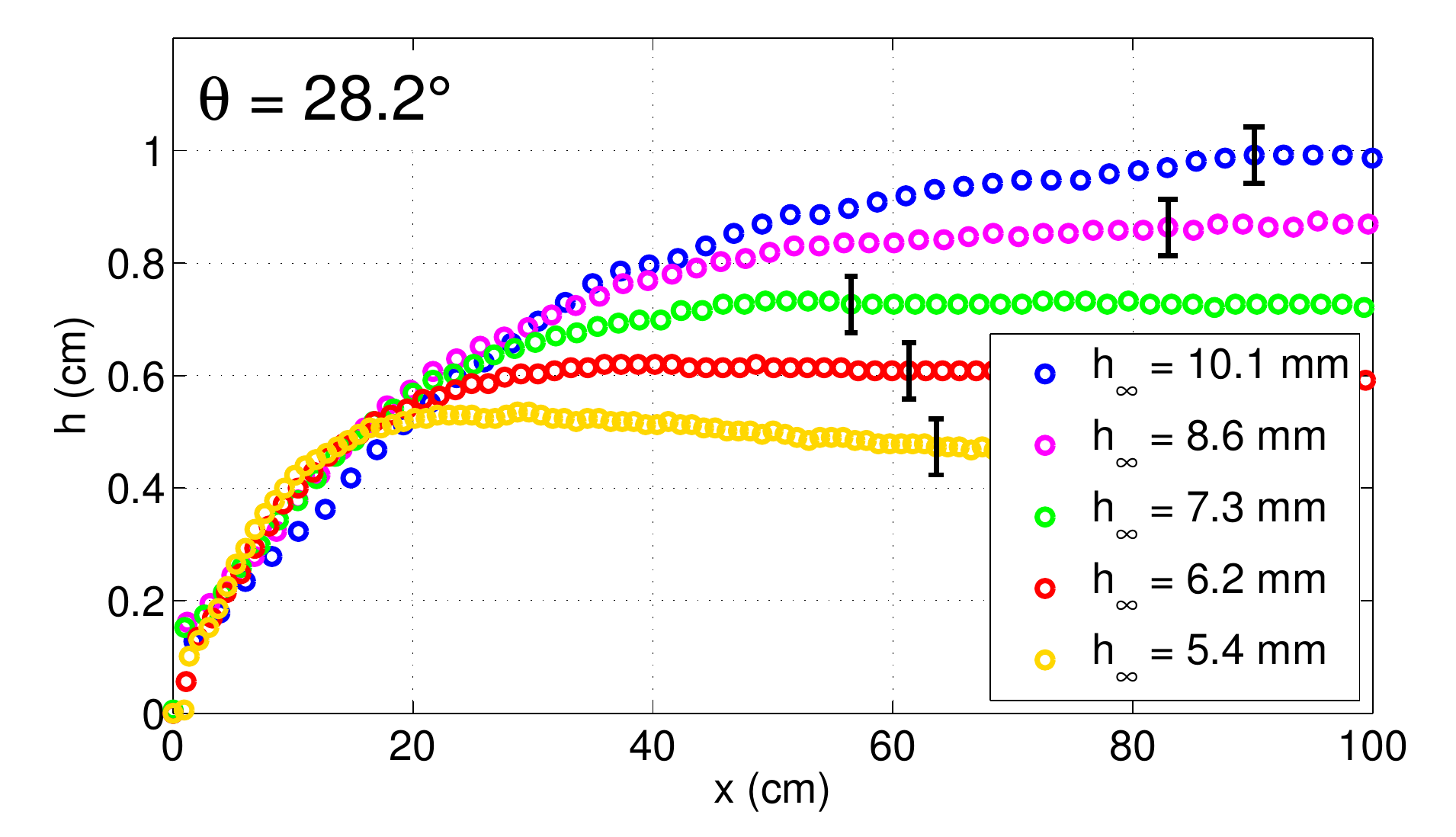} 
\includegraphics [width=80mm]{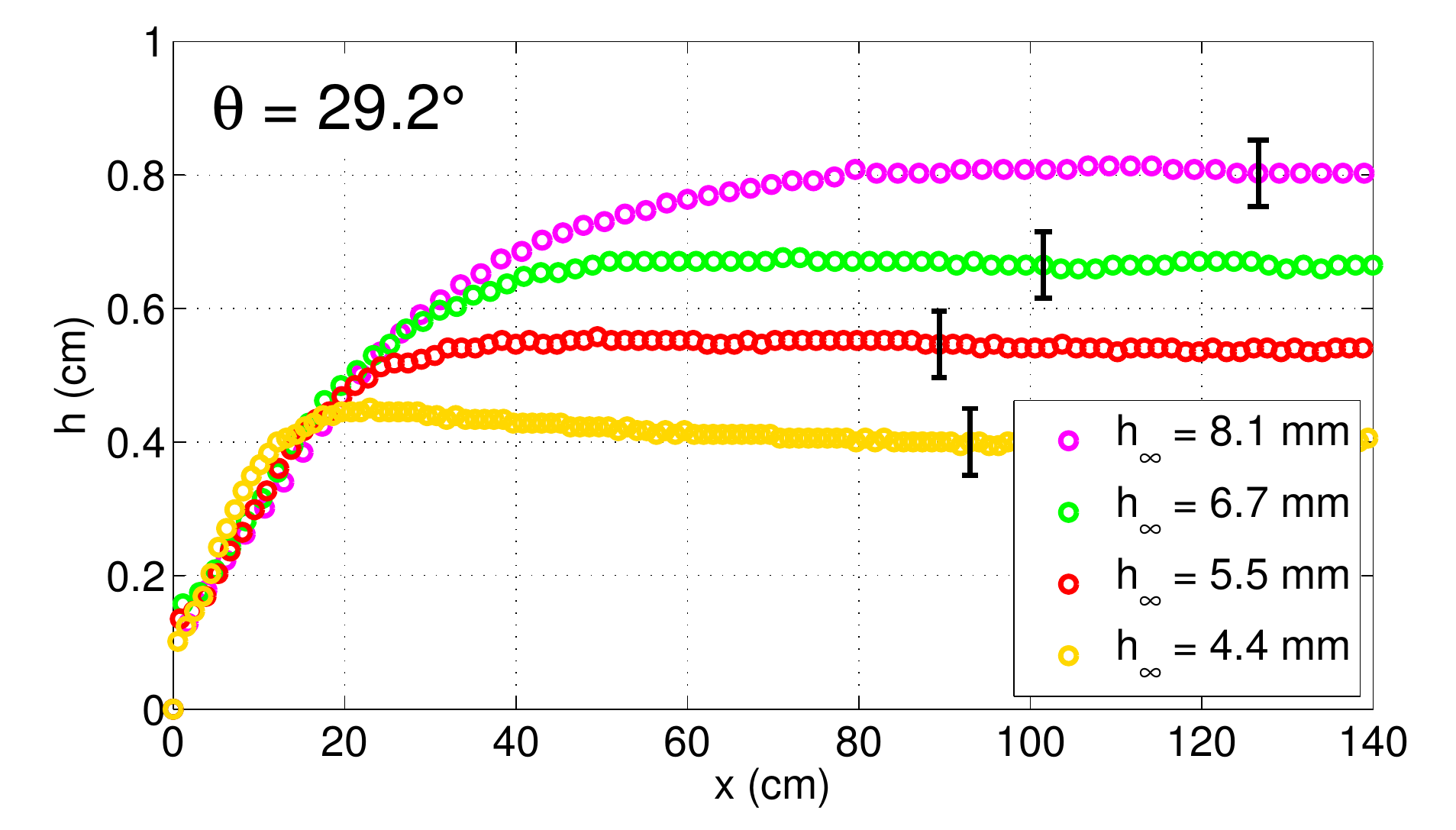} 
\captionof{figure}{Granular front profiles measured experimentally by transversal laser profilometry for different inclinations and different thicknesses  $h_\infty$, controlled by the aperture of the gate.}
   \label{data_brut}
\end{figure}

\begin{figure}[b] 
\center
\includegraphics [width=80mm]{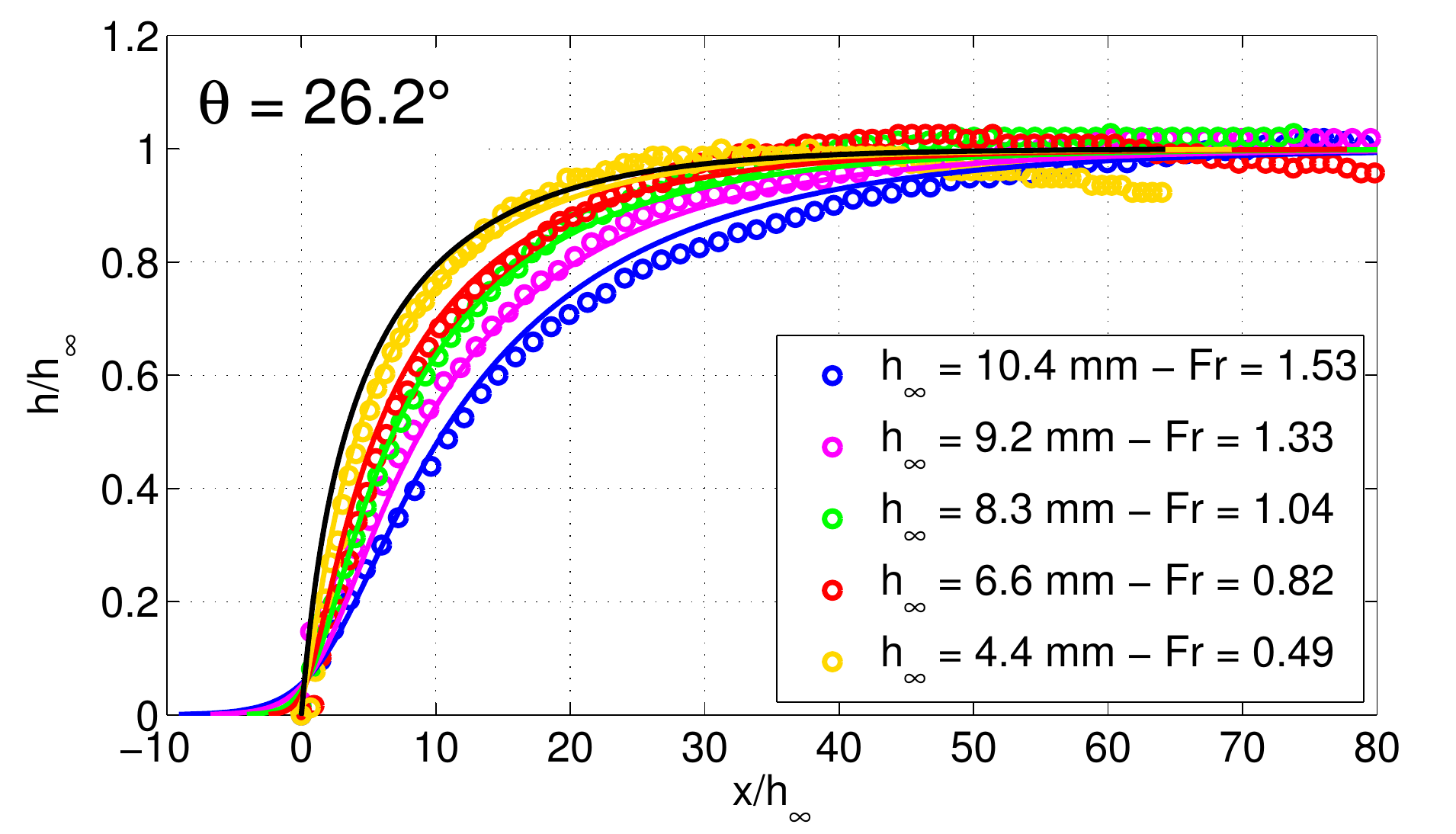} 
\includegraphics [width=80mm]{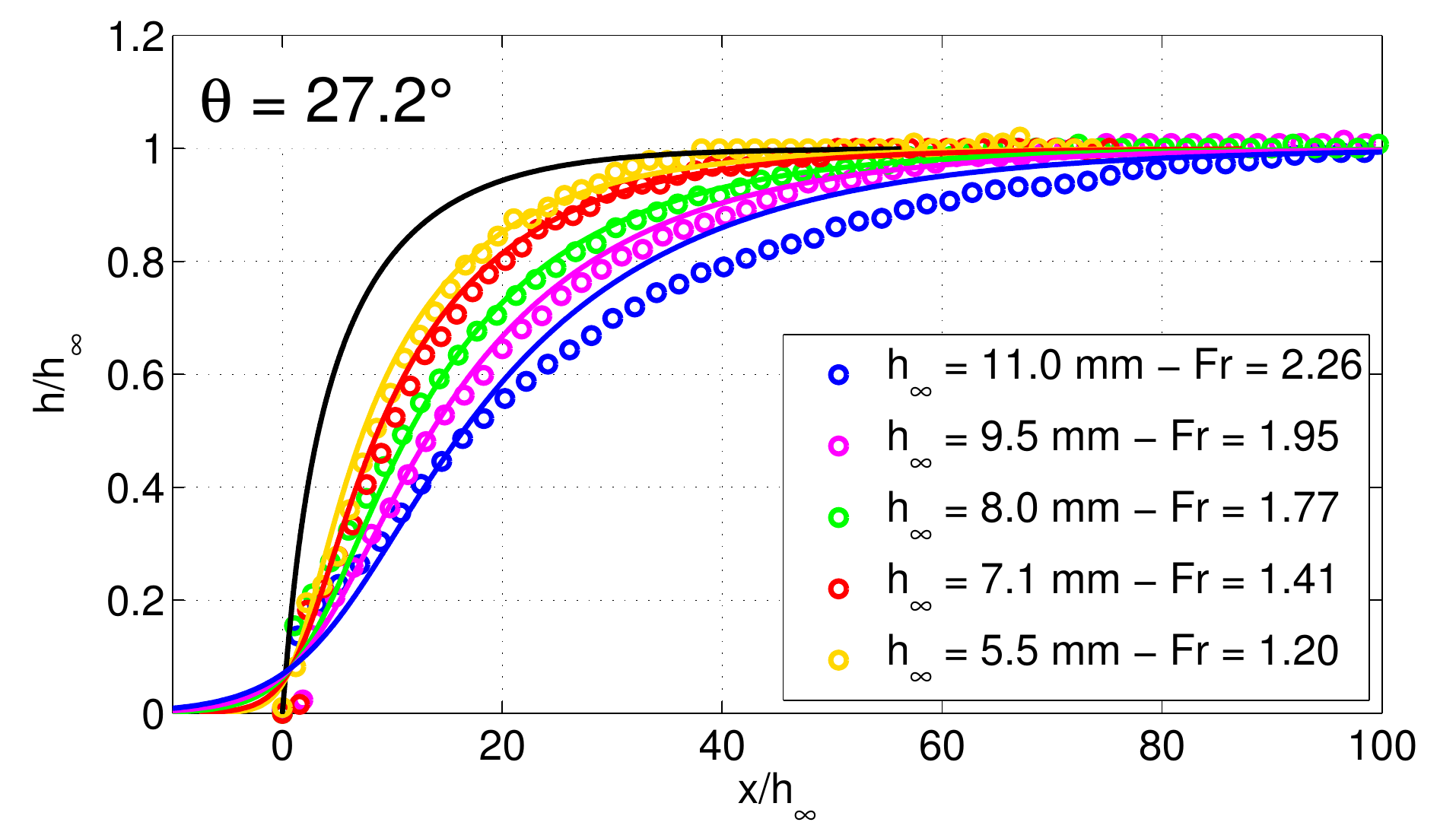} 
\includegraphics [width=80mm]{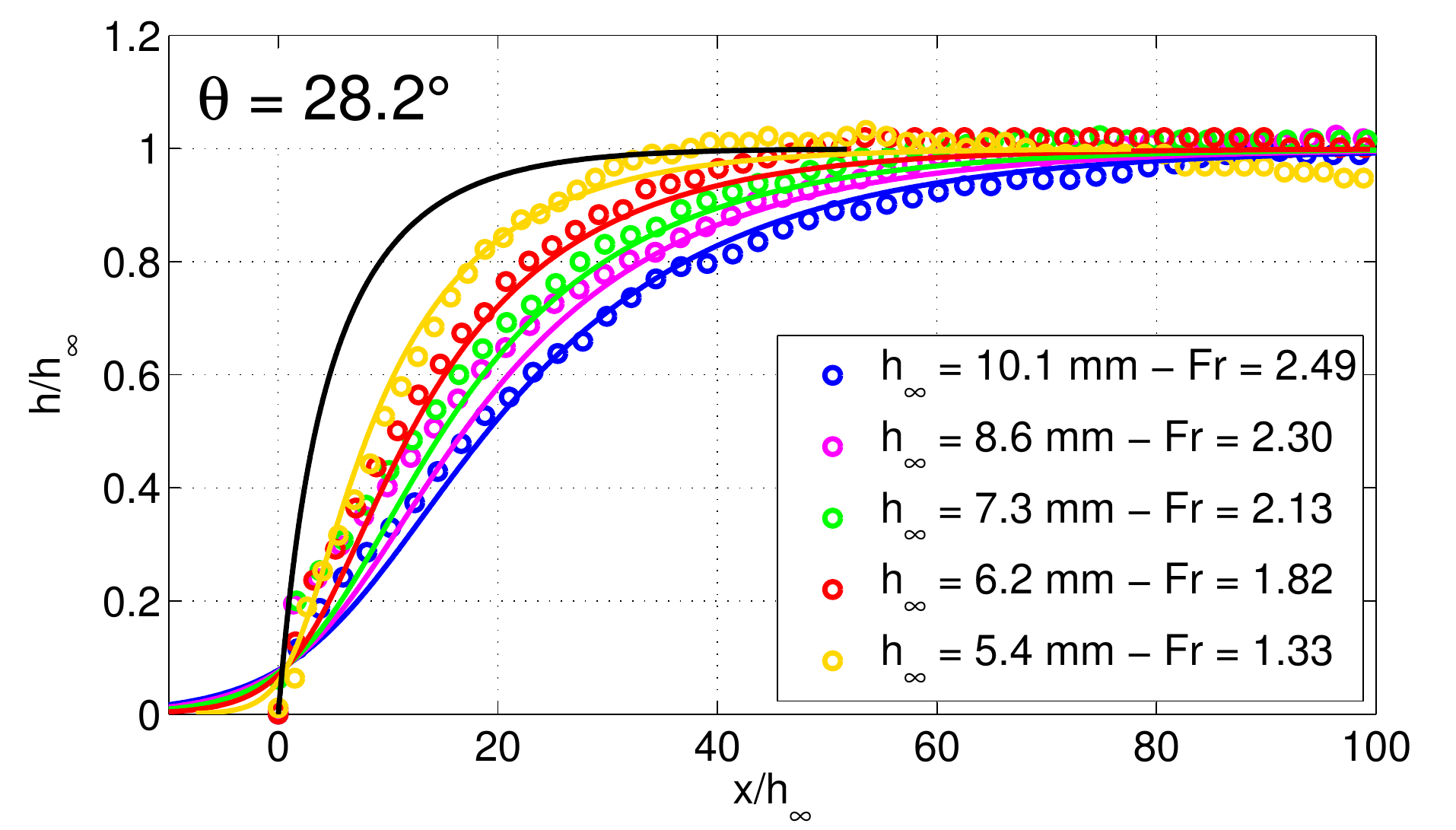} 
\includegraphics [width=80mm]{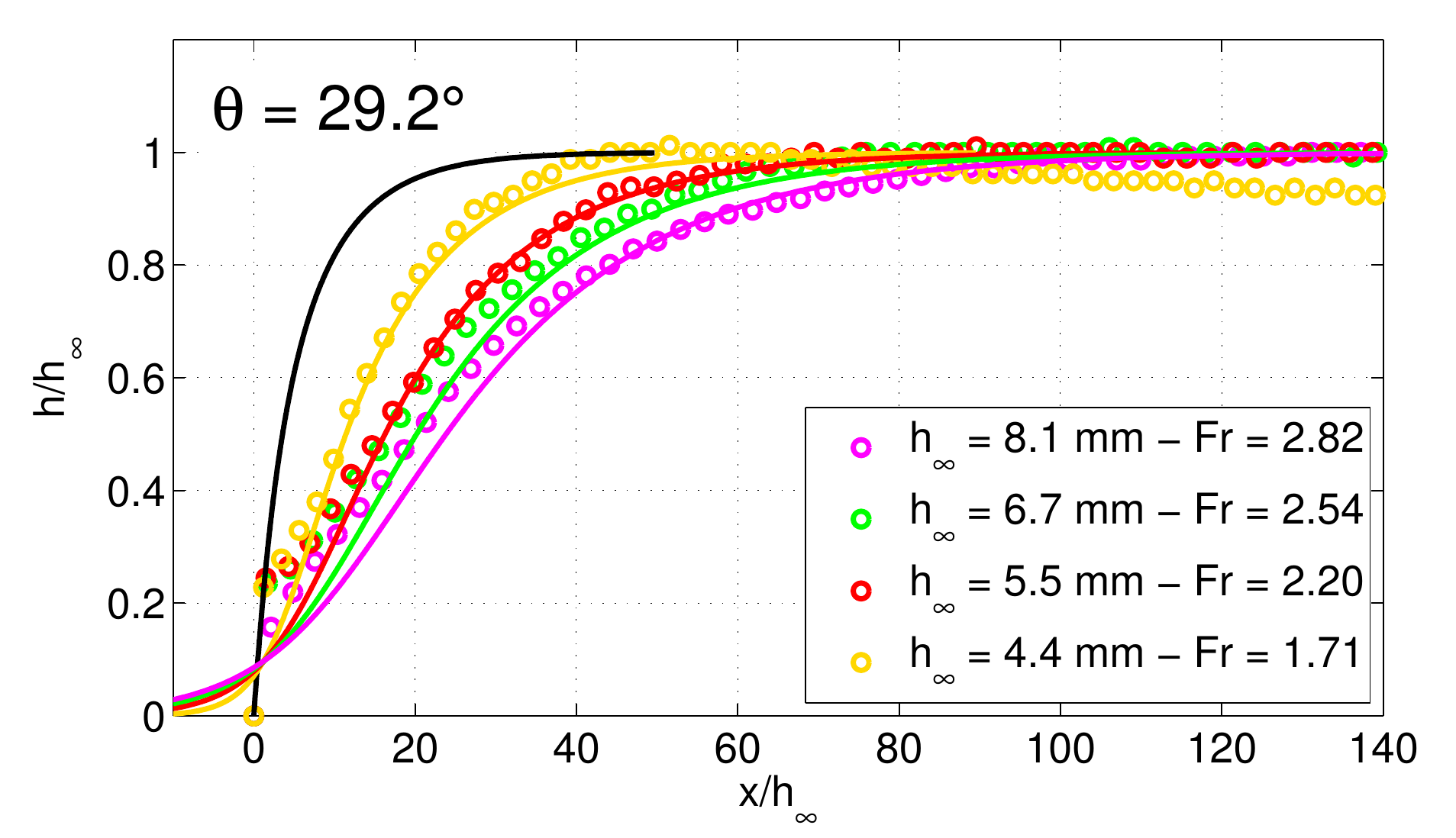} 
\captionof{figure}{Rescaled granular profiles: comparison between experiments and analytical predictions for different inclinations and different thicknesses $h_\infty$. Analytical solutions (colored lines) are calculated by using the thickness $h_\infty$ and the front velocity $u_0$ measured for each experimental front (colored circles) with a shape factor $\alpha = 5/4$. The analytical solution evaluated for $\alpha = 1$ is plotted in black line. Theoretical solutions are computed with rheology parameters $\mu_0 = 0.41$ and $\Delta\mu = 0.35$, determined by fitting our $h_{stop}$ data.}
    \label{data_rescaled}
\end{figure} 

\subsection{Second case: Inertial effect at higher Froude  numbers}

Now we consider the case of granular flows at larger Froude numbers, or more generally when  $(\alpha -1)Fr^2 \sim 1$. The inertial term cannot be neglected anymore in Eq. (\ref{St_Venant_qdm}) and Eq. (\ref{eq_gene_Fr}). This term adds a dependence of the front profile on the Froude number $Fr$ and the velocity profile through the value of $\alpha$.

We have realized new experiments with the set-up described in Sec. II,  allowing us to explore a more important range of velocity, from $10$ to $80$ cm/s (from $Fr=0.5$ to $Fr=3$) and to study the effect of inertia.  Some results of front profiles are presented in Fig. \ref{data_brut}. The thickness $h_\infty$ is measured with a precision of $\pm 0.5$ mm. After rescaling by $h_\infty$, the data of front profiles do not collapse and sort according to the front velocity, as shown in Fig \ref{data_rescaled}. The flattening of the front can be observed, as expected by the effect of the Froude number. By plotting the analytical solutions computed for $\alpha = 5/4 = 1.25$ (Bagnold-like profile), we have observed a good agreement between our experimental data and theoretical predictions. Moreover, the profile computed with $\alpha = 1$ is systematically above the other curves (see Fig. \ref{data_rescaled}), which proves that the hypothesis of a "plug flow" profile is  not adapted to describe the front of a granular flow on inclines at moderate or large Froude numbers.

\section{Discussion}

In this paper, we have derived an analytical solution for the front profile of a steady uniform flow on an incline from depth-averaged equations with the fractional frictional rheology $\mu(I)$ and with  a free shape factor $\alpha$ accounting for a non-constant vertical velocity profile. This model has been compared with experimental data, demonstrating the role of inertia and the influence of the free shape factor $\alpha$ on the front profile. 
In this part, we will discuss the influence of different parameters on the front profile. In a first time, we will focus on the choice of the velocity profile used to compute the analytical solutions and in a second time, we will analyze the effect of the rheology on our model and the consequences.

\subsection{Influence of the velocity profile}

In this work, we have shown the importance of the vertical velocity profile  in order to describe finely the shape of the front of a flowing granular layer. Contrary to many papers in the literature \cite{Savage_1989, Iverson_1997, Pouliquen_1999b, Mangeney_2003, Kerswell_2005, Gray_2014}, we have chosen a shape factor non equal to $1$. Indeed, in the case of a steady uniform granular flow on an inclined plane with a no-slip boundary condition at the bottom, we can demonstrate that the velocity profile should follow the Bagnold-like profile\cite{GDR_MiDi_2004}. Consequently, to compare the experimental results with the theoretical computations, we have supposed that this velocity profile was established in each point of the layer. Nevertheless, this hypothesis may be not satisfactory everywhere, in particular in the head of the front, which is greatly non-uniform and out of the theoretical Bagnold's limits. 

Even in the case of a steady uniform flow on an incline (far upstream to the front), some experimental and numerical data report a deviation from the Bagnold-like profile. Experimental results by Deboeuf \textit{et al.} \cite{Deboeuf_2006} report the ratio between the mean velocity and the surface velocity for steady-uniform granular flows of different thicknesses: this ratio increases from $1/2$ for thicknesses close to $h_{stop}$ to $3/5$ for thicker flows, that would correspond to linear and Bagnold-like profiles respectively (shape factors 
$\alpha$ equal to $4/3$ and $5/4$  respectively).  Discrete numerical simulations by Silbert \textit{et al.} \cite{Silbert_2001, Silbert_2003, GDR_MiDi_2004} show that the vertical velocity profile is a Bagnold-like profile in thick flows, whereas it is linear in thin flows. 
This raises the following question: what may explain the non-universality of a Bagnold-like profile for a steady and uniform flow on an incline? One possible reason would be the non generality of the no-slip boundary condition at the base. The role of the base roughness on the dynamics and on the boundary condition of the flow is not so clear as well. 

Moreover, by choosing an $\alpha$-value different of $1$, we have observed that the analytical solution presents an inflexion point near the head of the front, which leads to the creation of a precursor film. Experimental observations seem to invalidate this precursor film. For small Froude numbers, the front surface is well defined and does make  a finite contact angle with the plan. For higher Froude numbers, the precision of measurements is reduced due to splashes of grains downstream of the front. These splashes prevent a precise measurement of a contact angle but cannot be assimilated to a precursor layer. Again, all these results may indicate that the velocity profile is different in the head of the front from a Bagnold-like profile. 
Alternatively, to regularize this asymptotic behaviour, we could introduce a cut-off length that would correspond to the size of a few grains for instance, as it is  done in fluid mechanics \cite{Mahady_2015}. As mentioned by Hogg \& Pritchard\cite{Hogg&Pritchard_2004}, the definition of a non-constant shape factor $\alpha$ may also resolve this issue and lead to a best agreement between analytical solutions and experimental measurements near the head of the front. This method is commonly used in fluid mechanics where equations can admit a family of solutions for a family of velocity profiles \cite{Schlichting_1979}.

To finish, it may seem  irrelevant to compute shallow-water equations for granular flows on inclines with $\alpha=1$ at first sight, in particular with the knowledge of the Bagnold-like profile for the velocity. However, after writing here the equations for a free value of $\alpha$ in the case of the steady propagation without deformation of the granular front, it appears that this computation ($\alpha=1$) is equivalent to neglect inertia. Thus this work gives  a justification to this approximation. We may wonder to which extent this approximation can be extended to shallow granular flows in general? In other words, are there other configurations for which neglect inertia is equivalent to take $\alpha=1$? This would allow to extend studies from the literature done with shallow-water equations and $\alpha=1$ to the case of granular flows at small Froude numbers, whatever the vertical velocity profile is.

\subsection{Influence of the rheology parameters}

\begin{figure}[b] 
\center
\includegraphics [width=120mm]{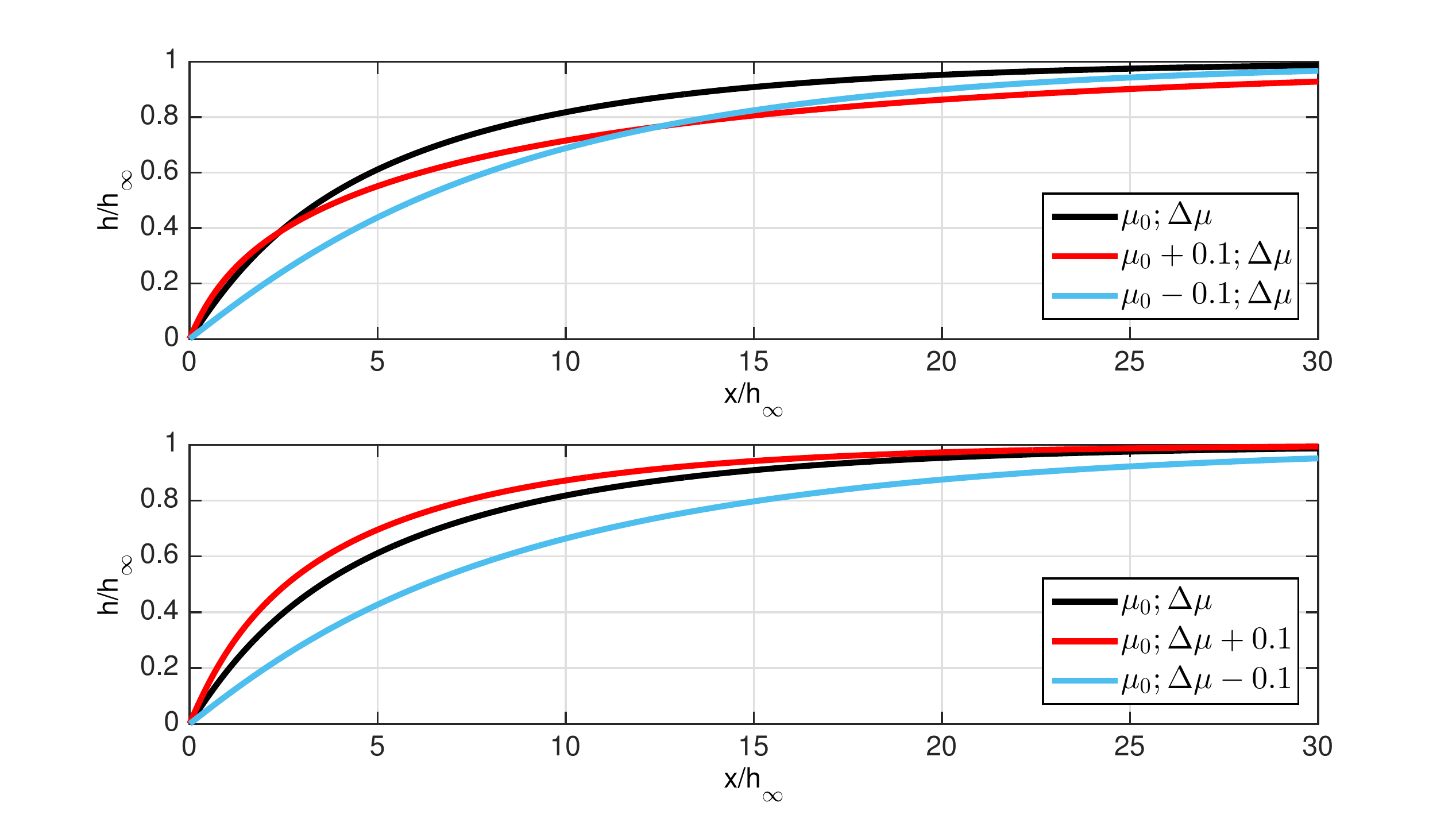} 
\captionof{figure}{Sensibility of the front profile to variations of each rheological parameter. Black curves are plotted for the inclination $\theta=29^\circ$ with $\mu_0$ = 0.41 and $\Delta\mu$ = 0.35, while other colored curves are for $\mu_0 \pm 0.1$ and $\Delta\mu \pm 0.1$ at constant $\Delta\mu$ and $\mu_0$ respectively.}
    \label{sensibility}
\end{figure}

Our analytical solution for the front profile (Eq. \ref{sol_anal_impl}) is written for the friction law expressed with the fractional expression as  $\mu_{0} + \Delta\mu/(1+I_{0}/I)$ (Eq. \ref{rheol_frac}), characterized by three free parameters $\mu_{0}$, $\Delta\mu$ and $I_{0}$. However, the rescaled profiles $h/h_{\infty}$ versus $x/h_{\infty}$ (Eq. \ref{sol_anal_impl}) do not depend on $I_0$: only two parameters - $\mu_0$ and $\Delta\mu$ - control the non-dimensionalized front shape. The sensibility on each parameter is evaluated by plotting front profiles for several values of $\mu_0$ and $\Delta\mu$ in Fig. \ref{sensibility}. Finally the value of $I_0$ only selects the steady thickness of the flow $h_\infty$.

For historical reasons (see below), the friction law for a given granular system is usually deduced from fitting $h_{stop}(\theta)$ and  $Fr(h/h_{stop})$ experimental data. 
Nevertheless, the range of measured  $h_{stop}$ data is restricted generally (between 1 and 10 grain diameters) and fits usually used are very sensitive to small values of $h_{stop}$. Consequently, the calibration of the rheology is sensitive to the precision and the error bar on each $h_{stop}$ point. In particular, an error corresponding to one size of grain can cause significant variations on the rheological parameters and modify the front morphology (see Fig. \ref{sensibility}).  
This sensitivity could be overtaken if the rheological parameters have physical interpretations (e.g., static and dynamic friction coefficients for $\mu_0$ and $\mu_0+\Delta\mu$). However, when experimental data of $h_{stop}(\theta)$ are fitted either by the fractional expression $\mu_{0} + \Delta\mu/(1+I_{0}/I)$ (Eq. \ref{rheol_frac}) or by the exponential expression $\mu_0 +\Delta\mu \exp(-I_0/I)$ as 2 examples, the values of friction parameters $\mu_0$ and $\mu_0+\Delta\mu$ are not the same, preventing to generalize any definition of these fit-dependent parameters. 
This raises the open question of a fine calibration of the frictional rheology from experimental data. 

\begin{figure}[t] 
\center
\includegraphics [width=80mm]{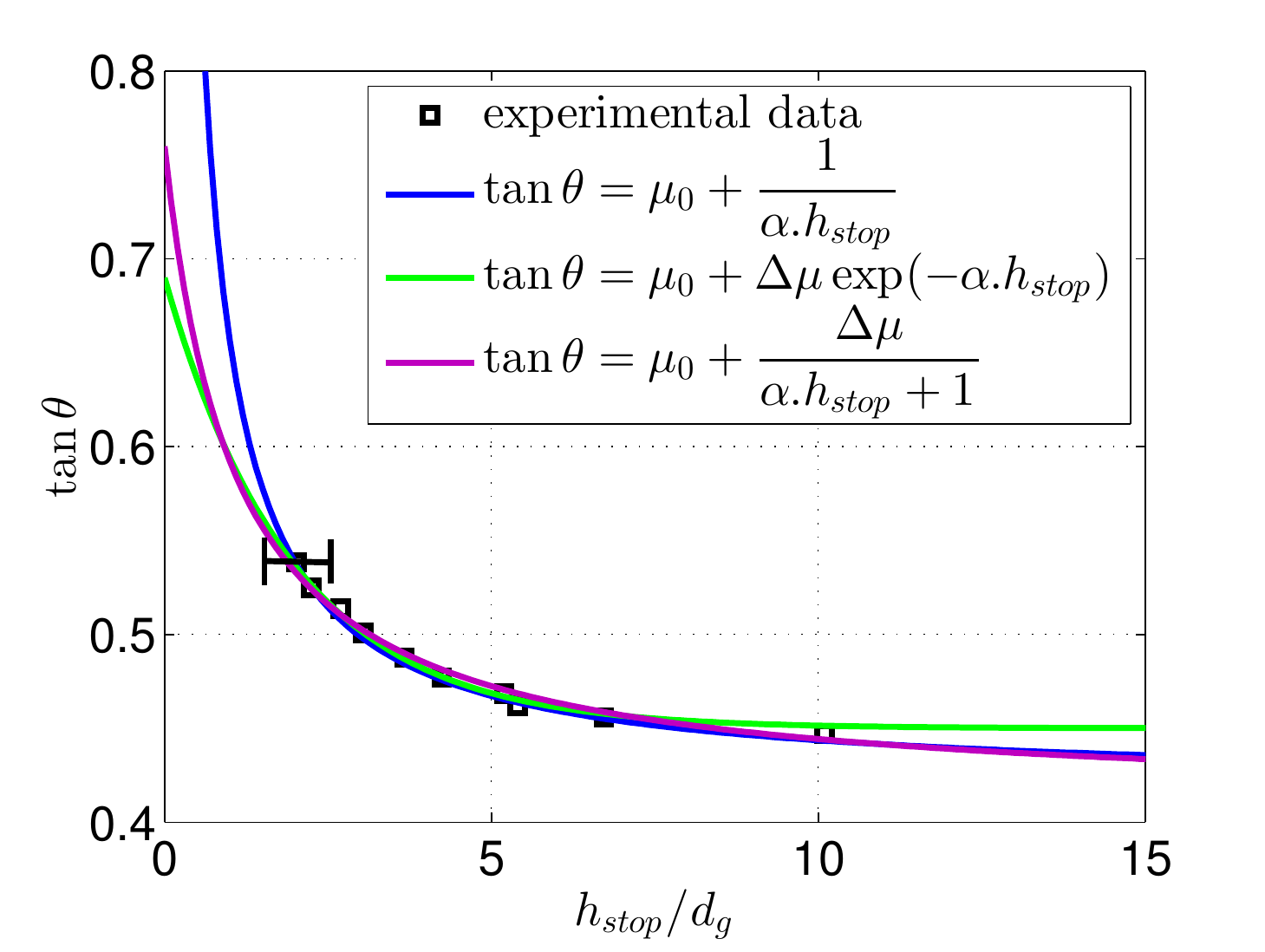} 
\includegraphics [width=80mm]{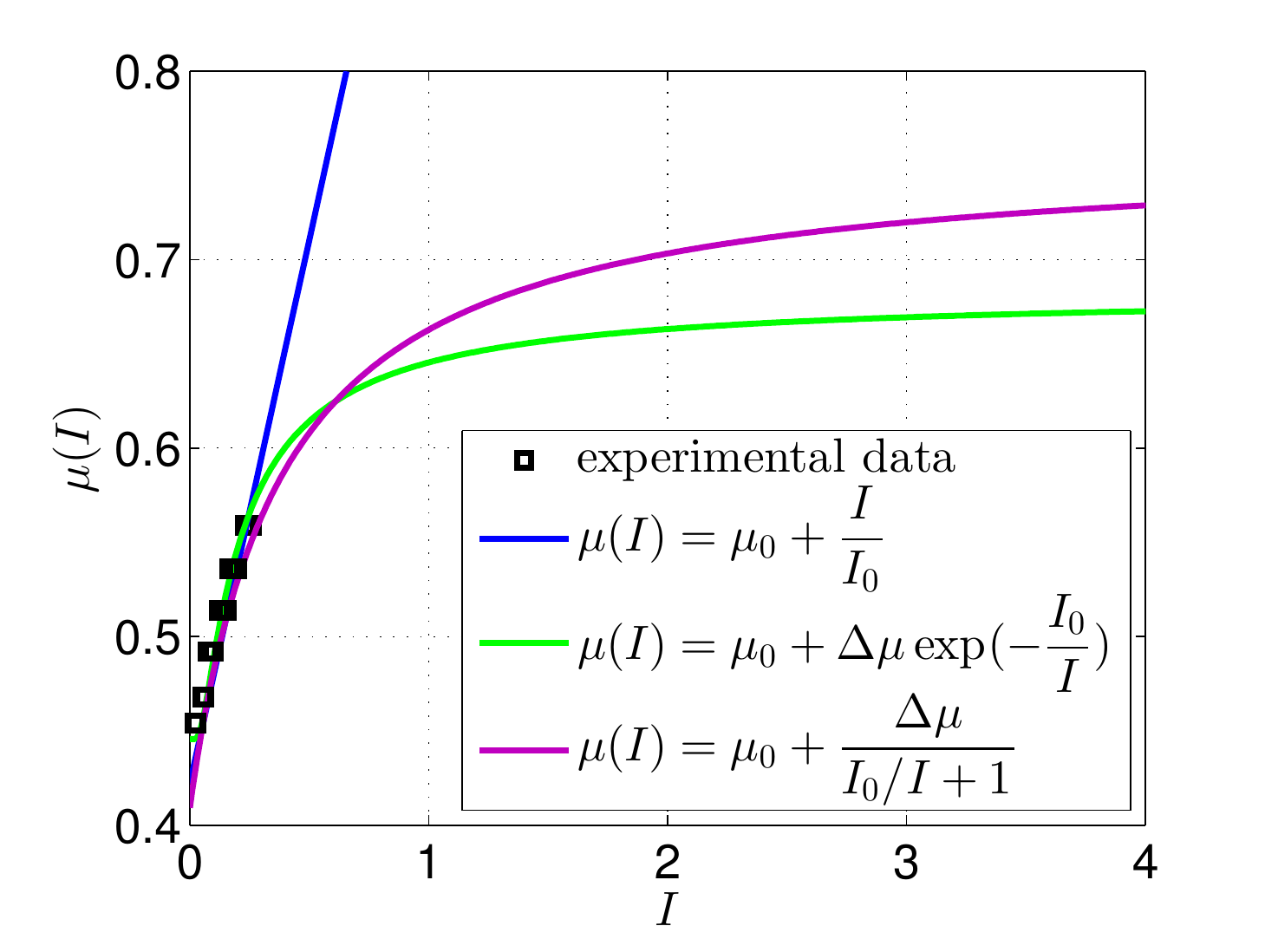} 
\captionof{figure}{Empirical determination of the rheological parameters: (left) Angular dependance of thickness $h_{stop}$ versus $\theta$ fitted by different expressions. (right) Local rheology $\mu$($I$) deduced from $h_{stop}$($\theta$) with different expressions. Linear $\mu(I) = \mu_0 + I/I_0$ with $\mu_0$ = 0.42 and $\I_0$ = 1.73.
Exponential $\mu(I) = \mu_0 + \Delta\mu\exp(-I_0/I)$ with $\mu_0$ = 0.45, $\Delta\mu$~=~0.24 and $I_0$ = 0.17.
Fractional $\mu(I) = \mu_0 + \Delta\mu/(1+I_0/I)$ with $\mu_0$ = 0.41, $\Delta\mu$~=~0.35 and $I_0$ = 0.38. }
    \label{rheology_hstop}
\end{figure}

Let us come back to the calibration of the friction law for a granular set-up. 
A major work, precursor of the friction law, was published by Pouliquen\cite{Pouliquen_1999a} reporting one relation between $h_{stop}$ and $\theta$ and another relation between $Fr$ and $h/h_{stop}$,  allowing him to write the basal friction coefficient from the parameters of these two relations. This indirect method  is usually used to determine the relation $\mu(I)$, especially for grains flowing on an incline. One paradox of this method is the use of $h_{stop}$ data, whereas the rheology $\mu(I)$ does not predict the existence of a deposit  or a threshold thickness  depending on the slope, but instead predicts the existence of one slope threshold. 
Another way of determining the expression of $\mu(I)$ would be to fit data of $\mu$ and $I$ without using the two previous relations, that would be a direct measurement of  $\mu(I)$. 

\begin{figure}[t] 
\center
\includegraphics [width=80mm]{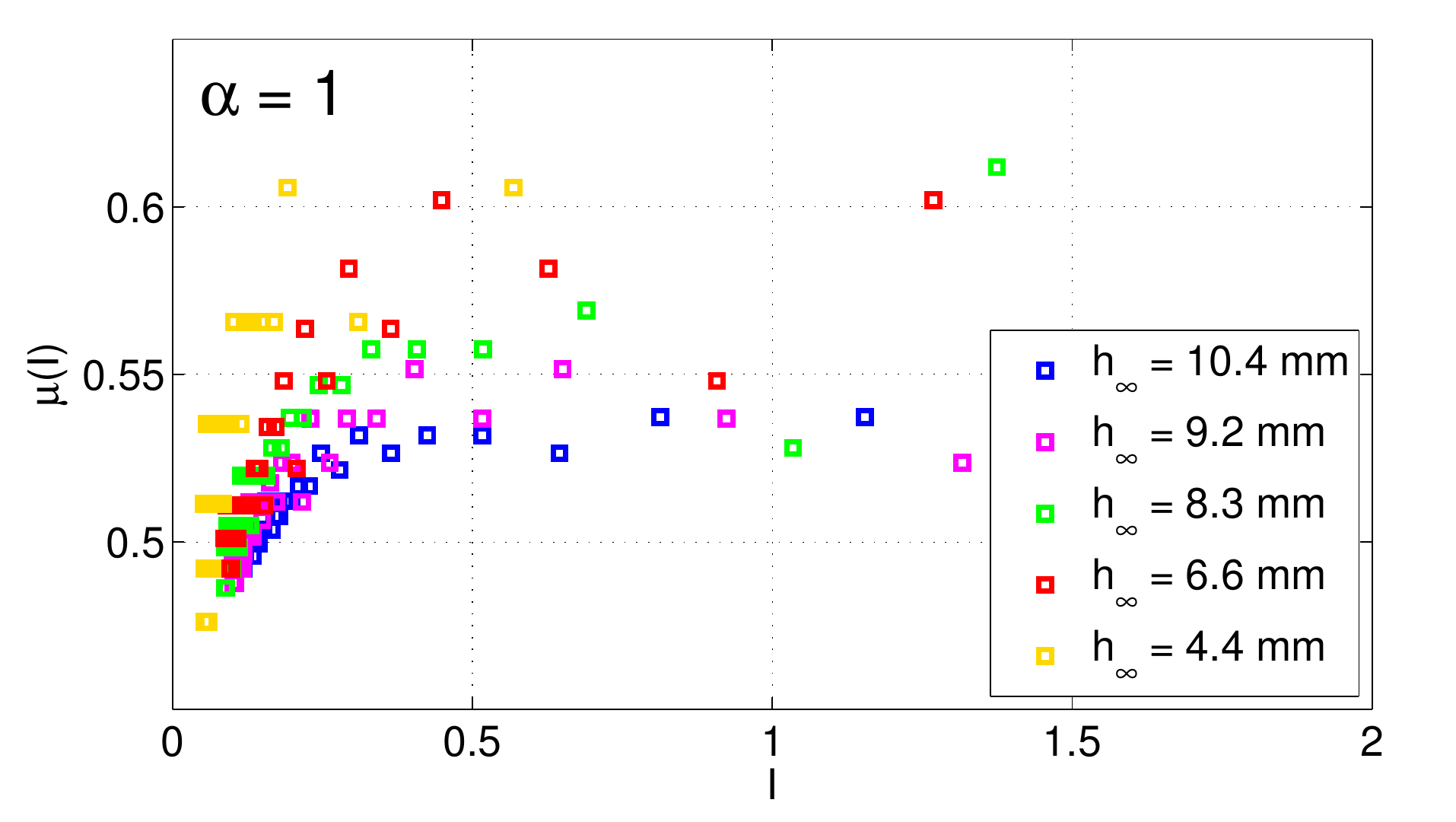} 
\includegraphics [width=80mm]{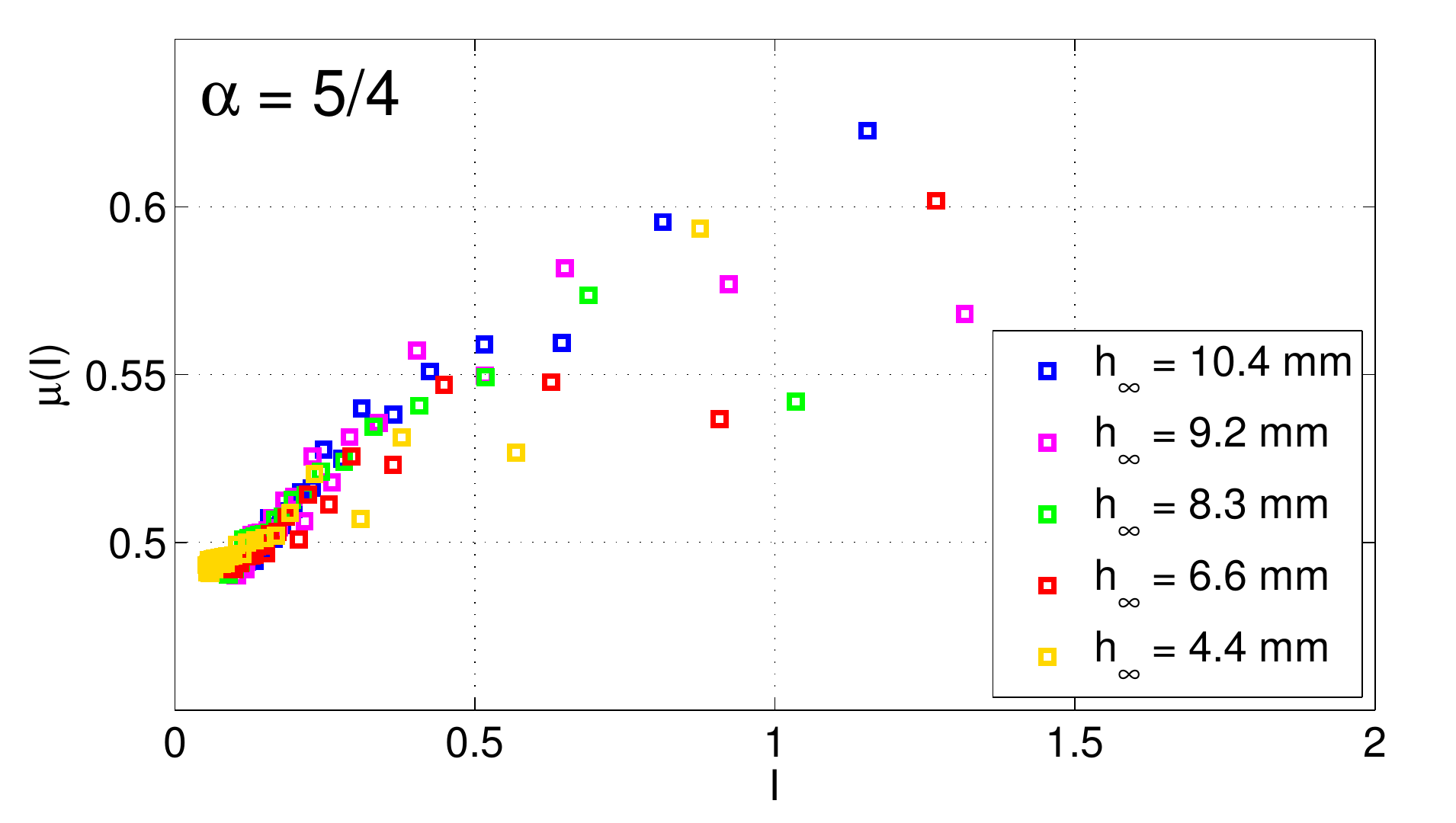} 
\captionof{figure}{Measurements of $\mu$($I$) from experimental front profiles at $\theta$ = 26.2$^\circ$ from Eqs.~(\ref{eq_simpl})  and~(\ref{eq_gene_Fr}) for $\alpha$ = 1 and $\alpha = 5/4$ respectively. The inertia number $I$ is computed from Eq.(\ref{defI}) with $\phi=0.6$ for the solid fraction. } 
    \label{rheology_front}
\end{figure}

To date there is nor consensus neither theoretical arguments leading to one expression for the friction law. Instead, we find in the literature 3 different functions:  
\begin{equation} 
\mu_{0} + \frac{\Delta\mu}{I_{0}/I + 1}, \ 
\mu_0 +\Delta\mu \exp(-I_0/I), \ 
\mu_0+I/I_0. 
\end{equation} 
In Fig. \ref{rheology_hstop} we show fits of $h_{stop}(\theta)$ with these different expressions and the deduced relations for $\mu(I)$ compared to the experimental data coming from steady uniform flows. By doing this, we can note that the range of $I$-values experimentally explored is not wide ($0.1<I<0.5$). We understand better that extending the rheology $\mu(I)$ from steady uniform flows to unsteady non-uniform flows was challenging for at least two reasons:  because of the introduction of unsteady and non-uniform terms in mass and momentum equations and because the values of inertia numbers may be outside the range of measurements of $I$ used for calibration. 
An alternative would be to use experimental measurements of $\mu(I)$ on a wider range of $I$ and/or from unsteady or non-uniform configurations. Is it possible from data of front profiles by using Eq. (\ref{eq_gene_Fr}) $\mu(I) = \tan\theta - \left[ (\alpha - 1)Fr^{2} h_\infty / h + 1 \right] d h / d \xi$, which can be written for small Froude numbers (or for $\alpha=1$) as: $\mu(I) \approx \tan\theta - d h/d \xi$  (Eq. (\ref{eq_simpl})). To this aim, we see that it is crucial to know $\alpha$ everywhere in the front. 
Fig. \ref{rheology_front} shows data points from a set of experiments realized at the same slope assuming $\alpha=1$ and $\alpha=5/4$. For $\alpha = 5/4$, data collapse for several thicknesses whereas they do not for $\alpha = 1$. 

\section{Conclusion}

We have proposed a new theoretical model to describe the shape of a granular front of a steady uniform flow on an incline. This model results from the shallow water Saint-Venant equations in 1D by considering a general velocity profile instead of a plug flow in the granular layer. By using a Bagnold-like velocity profile, we have demonstrated that inertial terms generate a front flattening when the Froude number associated to the flow increases. 

Our model was firstly compared to experimental data coming from Pouliquen \cite{Pouliquen_1999b}. In this case, the inertial effect is negligible. By rescaling experimental fronts at a given slope, data collapse onto one single profile. By taking into account the inertial corrections,  front profiles are roughly the same. We have provided new experimental results at higher Froude numbers that highlight the effect of inertia, which was neglected in previous models \cite{Savage_1989, Pouliquen_1999b}. A good agreement is found by comparing experimental data to theoretical predictions by assuming a Bagnold-like velocity profile established everywhere in the layer.   

These conclusions give reasons to perform new experimental investigations in order to determine the velocity field inside a granular front. Numerical discrete simulations can also provide interesting information to get a better understanding. Another approach would consist to investigate granular fronts with continuous numerical simulations \cite{Popinet_2003, Lagree_2011, Staron_2014}.

\textbf{ACKNOWLEDGMENTS}

The authors would like to thank O. Pouliquen for sharing his experimental data and for stimulating discussions, O. Dauchot, S. Popinet and C. Josserand for stimulating discussions and A. Lucquiaud for his preliminary work on the subject.

  \bibliographystyle{ieeetr}
  \bibliography{bibliography_steph} 
\end{document}